\documentclass[traditabstract]{aa}

\usepackage{graphicx}
\usepackage{txfonts}
\usepackage{natbib}
\usepackage{longtable,pdflscape}
\usepackage{booktabs}
\usepackage{multirow}
\usepackage[breaklinks, colorlinks, citecolor=blue]{hyperref}
\usepackage[flushleft]{threeparttable}
\def\arcs{\ifmmode {^{\scriptstyle\prime\prime}}
          \else $^{\scriptstyle\prime\prime}$\fi}
\def\parcm{\sa=.08em \sb=.03em
     \ifmmode \hbox{\rlap{.}\kern\sa}^{\scriptstyle\prime}\hbox{\kern-\sb}
     \else \rlap{.}\kern\sa$^{\scriptstyle\prime}$\kern-\sb\fi}
\def\arcm{\ifmmode {^{\scriptstyle\prime}}
          \else $^{\scriptstyle\prime}$\fi}
\def\parcs{\sa=.07em \sb=.03em
     \ifmmode \hbox{\rlap{.}}^{\scriptstyle\prime\kern -\sb\prime}\hbox{\kern -\sa}
     \else \rlap{.}$^{\scriptstyle\prime\kern -\sb\prime}$\kern -\sa\fi}

\def\apjl{ApJL}

\begin{document}

\title{RXCJ1111.6+4050 galaxy cluster: the observational evidence of a transitional fossil group}
\titlerunning{}

\author{R.~Barrena \inst{1,2} \and G. Chon \inst{3} \and H. B\"ohringer \inst{4,3} \and J. M\'endez-Abreu \inst{1,2} 
\and A. Ferragamo \inst{1,2,5}}
\institute{Instituto de Astrof\'{\i}sica de Canarias, C/V\'{\i}a L\'{a}ctea s/n, E-38205 La Laguna, Tenerife, Spain\\
\email{rbarrena@iac.es} 
\and
Universidad de La Laguna, Departamento de Astrof\'{i}sica, E-38206 La Laguna, Tenerife, Spain
\and
Universit\"ats-Sternwarte M\"unchen, Fakult\"at f\"ur Physik,
Ludwig-Maximilian-Universit\"at M\"unchen, Scheinerstr. 1, D-81679 M\"unchen,
Germany
\and
Max-Planck-Institut f\"ur extraterrestrische Physik, D-85748 Garching, Germany
\and
Dipartimento di Fisica, Sapienza Universit\`a di Roma, Piazzale Aldo Moro 5, I-00185 Roma, Italy
}

\date{Received ; accepted } 

\authorrunning{Barrena et al.}

\abstract{We present a detailed kinematical and dynamical study of the galaxy cluster RXCJ1111.6+4050 
(RXCJ1111), at $z = 0.0756$ using 104 new spectroscopic redshifts of galaxies observed at the Telescopio 
Nazionale Galileo and SDSS DR16 public archive. Our analysis is performed in a multiwavelength context in 
order to study and compare mainly optical and X-ray properties using XMM-Newton data. We find 
that RXCJ1111 is a 
galaxy cluster showing a velocity distribution with clear deviations from Gaussianity, that we are able to 
explain by the presence of a substructure within the cluster. The two cluster components show velocity 
dispersions of $644 \pm 56$ km s$^{-1}$ and $410 \pm 123$ km s$^{-1}$, which yield dynamical masses of 
M$_{200}$=$1.9 \pm 0.4 \times10^{14}$ M$_{\odot}$ and $0.6 \pm 0.4 \times 10^{14}$ M$_{\odot}$ for the main 
system and substructure, respectively. The 2D spatial distribution of galaxies and X-ray surface brightness 
of RXCJ1111 presents an elongation in the North-South direction. These observational facts, together with 
a gradient of 250-350 km s$^{-1}$ Mpc$^{-1}$ in the velocity field, following the NNE-SSE 
direction, suggest that the merger axis between the main system and substructure is slightly 
tilted with respect to the line-of-sight. The substructure 
is characterized by a magnitude gap $\Delta m_{12} \ge 1.8$, so it fits the "fossil-like" definition of a 
galaxy group. From the X-ray observations, we estimate a M$_{500,X}=1.68 \pm0.25 \times 10^{14}$ M$_{\odot}$, 
which is in good agreement with the dynamical masses when two galaxy components are considered separately. 
This suggests that the mass estimates obtained from X-ray and velocity dispersion are compatible even 
for non-relaxed clusters, at least when we are able to identify and separate galaxy clumps and 
derive masses by considering the virialized regions. We propose a 3D merging model and find that the fossil 
group is in an early phase of collision with the RXCJ1111 main cluster and placed at $\sim 8^\circ 
(\pm3^\circ)$ from line-of-sight. This merging model would explain the slight increase found in the T$_X$ 
with respect to what we would expect for relaxed clusters. Due to the presence of several brightest galaxies, 
after this collision, the substructure would presumably lose its fossil condition. Therefore, RXCJ1111 
represents the observational evidence that the fossil stage of a system can be temporary and transitional.}


\maketitle

\section{Introduction}
\label{sec:intro}

In recent decades, it has been well established that galaxy clusters are very dynamic structures in 
constant evolution that increase their mass in collision processes, either with low-mass systems or with 
massive clusters. Large mergers between high-mass clusters are among the most energetic events 
in the Universe. However, numerical simulations show that the accretion of small groups of galaxies is 
the main mechanism of evolution in clusters (e.g. \citealt{Ber09}; \citealt{McG09}). Today, much
observational evidence supports this hypothesis, such as the optical detection of substructures in 
the galaxy member distribution, inhomogeneities in the gas distribution by the study of X-ray and 
radio diffuse emission, or even the presence of complex dark matter halos in weak lensing data 
(\citealt{Fer02}; \citealt{Mar16}). In fact, studying mergers in low-mass systems is much more 
challenging than that in massive clusters. This requires hundreds of spectroscopic redshifts per 
cluster and X-ray observations with long exposures to obtain data with sufficient signal-to-noise 
from the weak diffuse emission.

The study of collisions involving small galaxy groups, and in particular the dynamical and kinematical 
properties, is a fundamental step in understanding the cluster evolution. One important question still
unclear is to understand the nature of fossil groups (hereafter FGs) and how these peculiar structures 
are formed and evolve. Fossil systems (groups and galaxy clusters) are dominated by a single luminous 
elliptical galaxy, similar to brightest cluster galaxies (hereafter BCGs) or even to cD galaxies, at the 
centre of the extended X-ray emission. Today, there exist several scenarios to explain the evolutionary 
picture in which FGs became fossils in the early Universe. One proposes that FGs grow through minor 
mergers alone, only accreting a few galaxies at $z \ge 1$, leaving FGs enough time to merge galaxies 
in one very massive and luminous object (\citealt{Pon94}; \citealt{Don05}). Another scenario 
proposes that FGs could be a transitional status for some systems. In this sense, fossil systems could
become non-fossil ones in the end due to the accretion of nearby galaxy systems, or even FGs could be 
swallowed by other more massive systems \citep{vBB08}. An alternative
evolutionary scenario suggests that FGs could be formed in the very early Universe but with a primordial
deficiency of mid- and low-luminous galaxies \citep{Mul99}.

RXCJ1111.6+4050 (hereafter RXCJ1111) was identified by G. O. Abell in 1958 and catalogued as 
Abell 1190 \citep{Abe58}, a galaxy cluster with richness R=2. RXCJ1111 was also detected in the ROSAT 
All Sky Survey (NORAS; \citealt{Boh00}) and designated as J1111.6+4050 as part of the MCXC catalogue 
\citep{Pif11}. And very recently, this cluster has also been observed using XMM-Newton X-ray 
satellite under the CHEX-MATE Heritage programme \citep{Arn21}. In optical, this cluster was also 
detected in the SDSS-DR6 and SDSS-DR8 photometric samples as a galaxy overdensity using the redMaPPer 
algorithm (\citealt{Wen09}, \citealt{Ryk16}). It has also been covered by the Pan-STARRS1 footprint 
and the Legacy Surveys DR9 images. In addition RXCJ1111 has been identified through its Sunyaev-Zeldovich 
(SZ) signal in the first and second Planck cluster catalogues and named as PSZ1 G172.64+65.29 
\citep{PC29} and PSZ2 G172.88+65.32 \citep{PC27}. The FIRST VLA Survey also covered this target 
region \citep{Bec95} in the 300 square degree initial observations.

In this work, we analyze the RXCJ1111 cluster of galaxies using optical, X-ray and radio data in order 
to disentangle the dynamical state and the main physical properties of this system which shows clear 
signs of substructure. Intra-cluster medium (ICM) and galaxy component react with different time scales 
to cluster evolution showing many observational effects at numerous spectral frequencies. Thus, 
multi-wavelength analysis of galaxy clusters is the ideal approach to investigate merging processes. 
Our main aim is to find a coherent dynamical scenario in agreement with the effects observed in 
different wavelengths. Only the combination of spectroscopic information, X-ray and radio observation 
will help us to obtain a satisfactory answer for questions such as, is this a merging cluster?, in 
that case, are we looking at pre- or a post-merger phase?, what kind of structures are involved?, are 
X-ray and dynamic mass estimate in good agreement?, ... Here, we will explore the dynamics of RXCJ1111 
with the aim of answering all these questions.

This paper is organized as follows. In Sect. \ref{sec:optical_obs} we describe the new spectroscopic 
observations as well as the X-ray data. We analyse the optical and galaxy properties of RXCJ1111 in Sect. 
\ref{sec:optical_ana} and \ref{sec:BCG_membership}, and compare them with X-ray properties in Sect. 
\ref{sec:Xprop}. In Sect. \ref{sec:mass} we present the main dynamical features and propose a plausible 
3D merging model for the cluster in Sect. \ref{sec:merging}. We conclude this paper summarising our 
results in Sect. \ref{sec:conclusions}.

In this paper, we assumed a flat cosmology with $\Omega_m=0.3$, $\Omega_\Lambda=0.7$ 
and H$_0=70$ $h_{70}$ km s$^{-1}$ kpc$^{-1}$. Under this cosmology, 1 arcmin corresponds to 87 
$h_{70}^{-1}$ kpc at the redshift of RXCJ1111 ($z = 0.0756$).

\section{Data sample}
\label{sec:optical_obs}

\subsection{Optical spectroscopy}
\label{sec:spectroscopy}

Despite RXCJ1111 having been observed in radio frequencies, X-ray and several broad-band photometric data, 
the spectroscopic information in the literature and databases is relatively poor. We selected 43 
spectroscopic redshifts from the SDSS-DR16 database within a region of $15'$ radius with respect to the 
centre of the cluster, which is too sparse a sample to investigate this low redshift cluster. 
Therefore, in June 2020 we carried out spectroscopic observations at the 3.5m Telescopio Nazionale 
Galileo (TNG) telescope at Roque de los Muchachos Observatory.

One of the most used techniques to obtain a large number of galaxy redshifts in a limited field is 
multi-object spectroscopic (MOS) observations. In June 2020, we carried out MOS observations
of RXCJ1111 covering a region of about $17^\prime\times17^\prime$. We mapped this region with 5 MOS
masks including 198 slitlets. The masks were designed in order to avoid overlaps with the SDSS 
redshift sample and maximise the number of new redshifts. We used the 3.5m TNG telescope and 
its spectrograph DOLORES. The instrumental set-up was used with the LR-B grism\footnote{See http://www.tng.iac.es/instruments/lrs} 
and slits of $1.6\arcsec$ width, which offers a dispersion of 2.75 \AA $\ $ per pixel between 
370 and 800 nm of wavelength coverage. We acquire a single 1800 s exposure per mask.

\begin{figure*}[ht!]
\centering
\includegraphics[width=\textwidth]{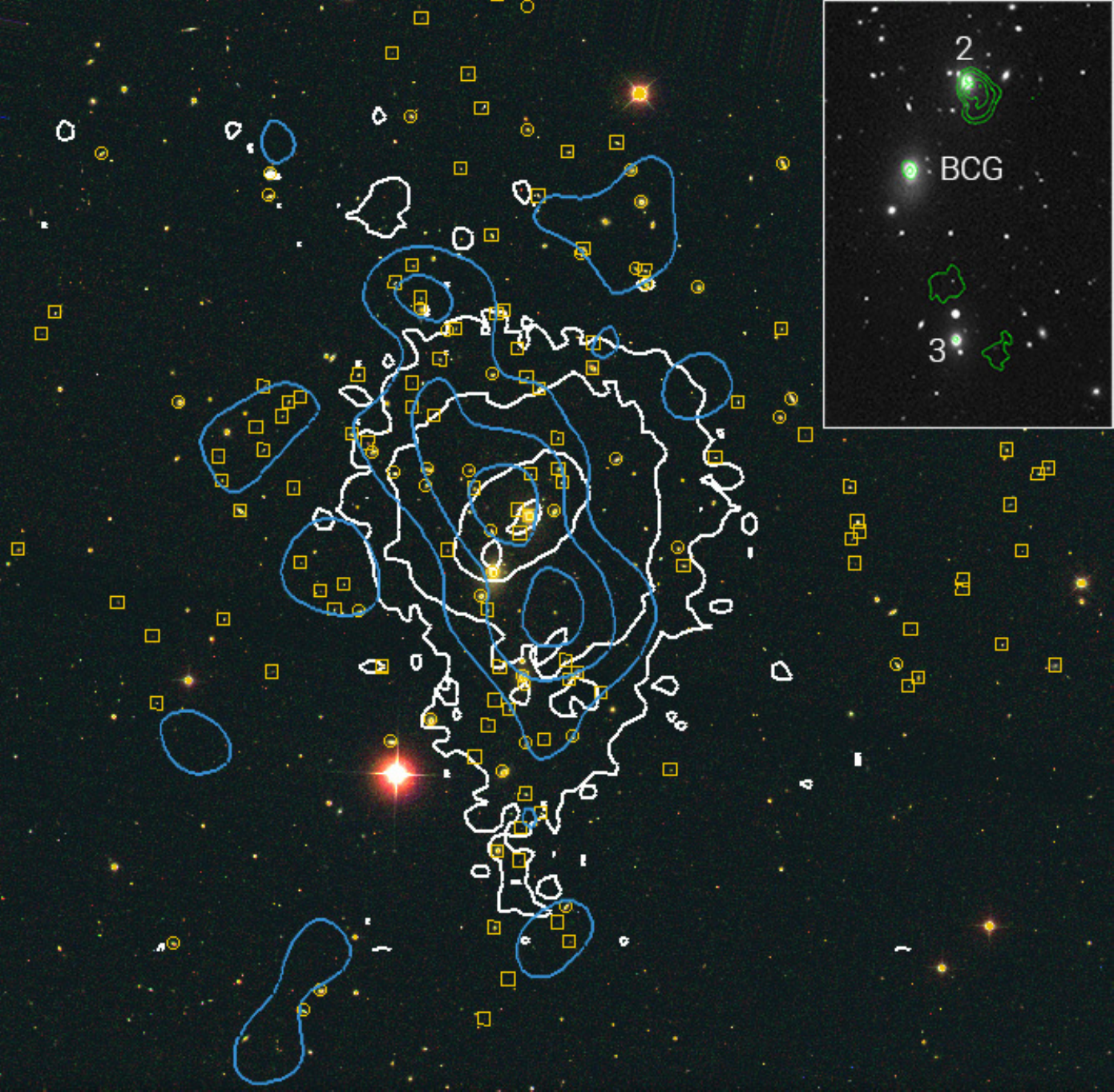}
\caption{RGB colour composite image obtained by combining $g'-$, $r'-$ and $i'-$band images of 
$23^\prime\times23^\prime$ field of view from Pan-Starrs1 public archive. Yellow squares and circles 
mark the galaxies observed in our spectroscopic MOS observations and SDSS-DR16 spectroscopic 
redshifts, respectively. Blue contours show the isodensity galaxy distribution of likely cluster 
members (see Sect. \ref{sec:spatial_distrib}). White contours correspond to X-ray surface brightness
after removing point sources using a pixel mask. Superimposed, in the upper right corner, the cluster
core is zoomed. Labels "BCG", "2" and "3" mark the brightest cluster galaxy (\textit{BCG}), the second and 
the third brightest galaxies (\textit{BCG2} and \textit{BCG3}), respectively. Green contours represent the diffuse 
radio emission observed with the Very Large Array (VLA) telescope.
North is upward and East to the left.}
\label{fig:contours}
\end{figure*}

The spectra were extracted using standard {\tt IRAF} packages and calibrated in wavelength
using Helium, Neon and Mercury lamps. The spectroscopic redshifts of galaxies were obtained
by correlating the scientific spectra with those assumed to be templates (from Kennicutt Spectrophotometric Atlas 
of Galaxies; \citealt{ken92}) using the technique by \citet{Ton79} and implemented as the task 
{\tt RVSAO.XCSAO} in {\tt IRAF} environment. This method detects and correlates the main
features present in the acquired spectra (i.e. Ca H and K doublet, H$_\delta$, G band, and MgI 
in absorption, and OII, OIII doublet, H$_\alpha$ and H$_\beta$ in emission) with 
that present in the template ones. We used five templates corresponding to different galaxy 
morphologies (Elliptical, Sa, Sb, Sc and Irr types). At the end of the process we obtained 
a radial velocity estimate and the corresponding correlation error for 109 galaxies in the
field of RXCJ1111. We added to this sample 43 redshifts retrieved from SDSS-DR16 spectroscopic
database. So, our spectroscopic sample (see table \ref{tab:catalog}) includes 152 
redshifts in a region of $17^\prime\times17^\prime$ (see Fig. \ref{fig:contours}). 
The full redshift sample presents a median SNR of 7 and a median error in $cz$ of 91 km s$^{-1}$, respectively. 
We detected 37 star forming galaxies, characterized by the presence of $[$OII$]$, $[$OIII$]$ 
and/or H$_\alpha$ emission lines with equivalent widths $>10$ \AA. 

Twelve target galaxies were observed in two different masks. These double redshift measurements allow
us to estimate realistic errors (including systematic ones) by comparing the two redshifts 
obtained with the {\tt XCSAO} correlation procedure. We find that both redshift
estimates are in agreement and their corresponding errors are similar. So, we confirm that 
{\tt XCSAO} provides, in this case, not only statistical errors, but also realistic 
uncertainties.

\begin{figure} 
\centering
\includegraphics[width=\columnwidth]{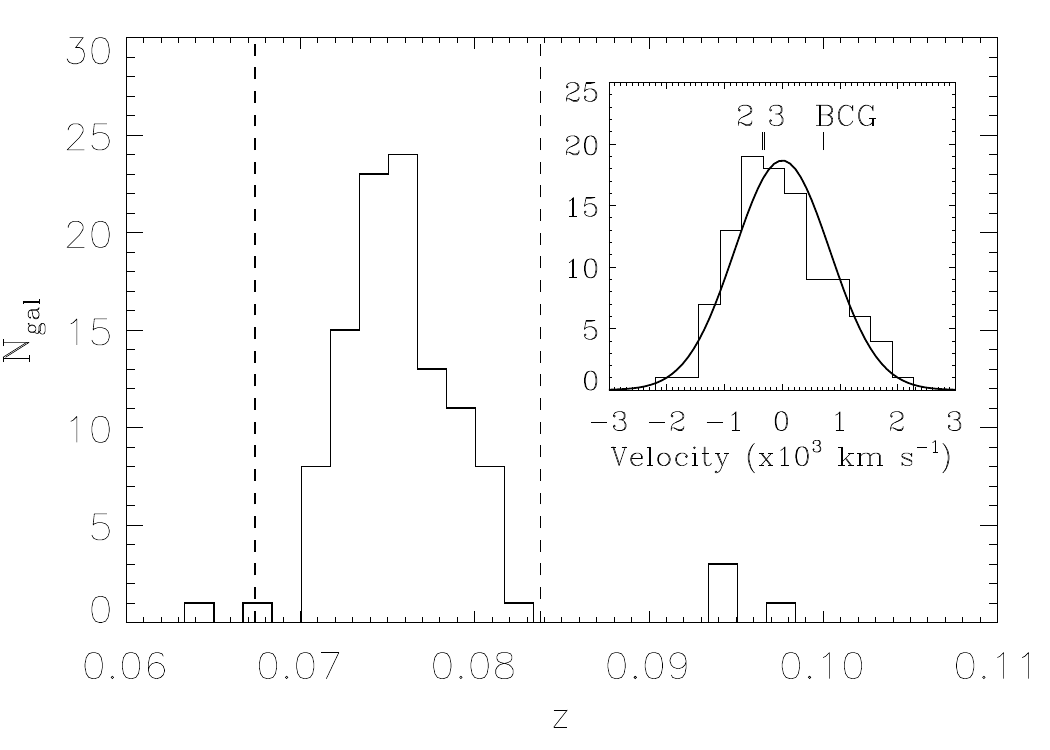}
\caption{Galaxy redshift distribution in the range $0.02<z<0.11$. Dashed vertical lines delimit the 
redshift range including 104 galaxy members assigned to RXCJ1111 according to 2.7$\sigma_\textrm{v}$
clipping. The inner plot shows the velocity distribution in the cluster rest frame. The black Gaussian 
curve represents the velocity reconstruction according to the biweight method and assuming all the galaxies
are part of a single system. The velocity corresponding to \textit{BCG}, the second and the third brightest 
galaxies are also marked with the labels '\textit{BCG}', '2' and '3', respectively.}
\label{fig:histogram}
\end{figure}

Table \ref{tab:catalog} lists the complete spectroscopic sample considered in this work (see 
also Fig. \ref{fig:contours}). Col. 1 lists an ID number (cluster members are marked), 
Cols. 2 and 3 show the J2000 equatorial coordinates of galaxies, Col. 4 the 
heliocentric radial velocity ($v=cz$) with their corresponding errors ($\Delta\textrm{v}$), and Cols. 5 and 6, the 
complementary $r'$ and $i'$ $dered$ magnitudes, respectively. The last column includes some comments 
regarding particular features of some galaxies. Fig. \ref{fig:histogram} shows the 
velocity distribution of galaxies around the cluster main redshift.

\subsection{Optical photometry}
\label{sec:photometry}

We also work with the SDSS DR16 photometric data in order to complement our spectroscopic MOS TNG observations. 
We consider the extinction-corrected $dered$ magnitudes $g'$ and $r'$, that assume the \cite{Sch98} reddening 
maps, within a circular region of $12^\prime$ radius. The mean depth (at $\sim$90\% completeness) 
of this photometric sample is $r'=21.5$, which is in agreement with SDSS DR12 
estimates\footnote{see https://www.sdss.org/dr12/imaging/other\_info/}. Comparing the photometric and
spectroscopic samples in that regions covered by the MOS masks, we find that the completeness of the 
spectroscopic sample is $\sim50$\% for galaxies down to magnitude $r'=18.5$. However, this completeness
increases up to $\sim60$\% for galaxies with magnitude $r' \le 19.5$ in an inner region of $5^\prime$ arcmin 
radius from the center of the cluster. The quality of the spectra allows us to obtain redshifts even for 
some faint galaxies with $r'>21$.

\subsection{X-ray data}
\label{sec:x-ray}

The XMM-Newton observation, with ID 0827031101, of the RXCJ1111 galaxy cluster, were obtained from 
the XMM-Newton data archive. This target was observed as part of the CHEX-MATE Cluster Heritage project 
(PIs: M. Arnaud and S. Ettori; A\&A 650, A104). We used SAS v20.0 to perform the X-ray imaging and 
spectroscopic data reduction  closely following the scheme described in~\citet{chon15}. After cleaning 
the data from times of X-ray flares, the usable exposure amounts to 34~ks for both MOS instruments 
and to 26~ks for pn.

\begin{figure} 
\centering
\includegraphics[width=\columnwidth]{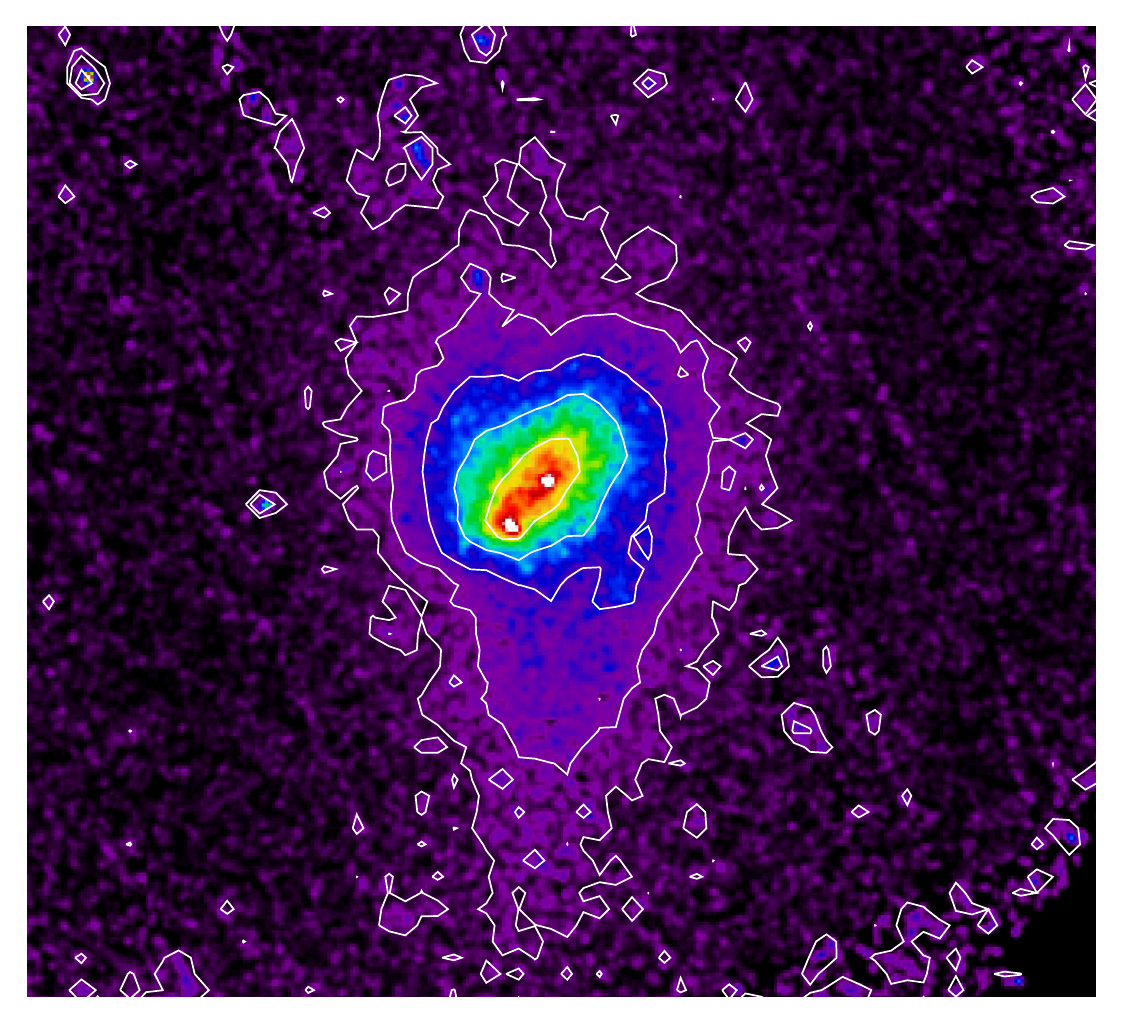}
\caption{XMM-Newton image of the cluster RXCJ1111.6+4050 in the 0.5 to 2 keV energy band.
The size of the image is 21.5 (width) by 20 (height) arcmin. North is upward and East is to the left.}
\label{fig:r1111_image}
\end{figure}

We removed point sources and the background-subtracted and exposure-corrected images from all three 
detectors were combined in the 0.5 to 2 keV band, which is shown in Fig.~\ref{fig:r1111_image}. 
RXCJ1111 has on intermediate scales a round appearance, but it shows a bright extension to the south 
and seems to be embedded in a larger north-south filament with a length of about 1.8~Mpc as far as 
it can be traced in X-rays. The centre is disturbed with a bar-like feature oriented in 
northwest-southeast direction with two small X-ray peaks inside.

For the spectral analysis, the contribution from the particle background was removed by rescaling the 
filter wheel closed (FWC) spectrum to the spectrum of the corner events of the observation. We 
considered three X-ray background components representing the unresolved point sources, the Local Hot 
Bubble and a cool absorbed thermal model when fitting an APEC cluster model to the spectroscopic data 
in XSPEC.

\section{The RXCJ1111 optical properties}
\label{sec:optical_ana}

\subsection{Member selection and global properties}
\label{sec:members_global}

\begin{figure} 
\centering
\includegraphics[width=\columnwidth,height=9cm]{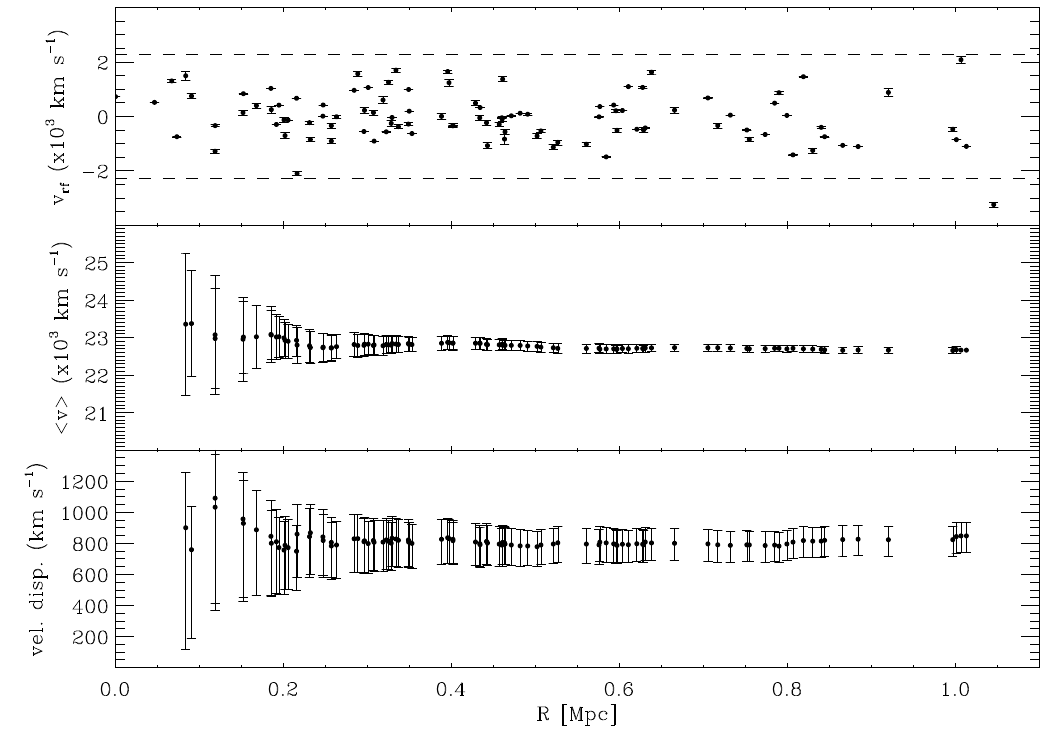}
\caption{Top panel: Measured LOS velocity, in the cluster rest frame, of the 104 galaxy members versus 
projected distance to the centre. The cluster centre is assumed to be the position of the \textit{BCG}. Middle 
and bottom panels: Integral mean velocity and LOS velocity dispersion, also in the cluster rest frame, 
shown as radial profiles with respect to the cluster centre. These values are computed by considering 
all galaxies enclosed in that radius. The first value computed is estimated from the first five galaxies 
closest to the centre. The error bars are at the 68\% c.l.}
\label{fig:3plots}
\end{figure}

In order to analyse the internal dynamics of RXCJ1111, it is essential to carry out a good selection of 
member galaxies. For this purpose, one of the most suitable methods is the one that uses the velocity 
"caustics", which is related to the escape velocity from the cluster (\citealt{Dia05}; \citealt{Lem09})
allowing  to separate cluster members from foreground and background galaxies. However, this method works
well with large spectroscopic samples, typically with more than 300 redshifts, and we do not find reliable 
results applying this procedure to our sample. Thus, we use a similar but more simple technique, based on 
the galaxy position in the projected ($r$, $cz$) space, where $r$ is the projected cluster-centric 
distance and $cz$ is the galaxy line-of-sight (LOS) velocity (see Fig. \ref{fig:3plots}, top panel). We 
apply an iterative 2.7$\sigma_\textrm{v}$ clipping in the $cz$ coordinate, considering a radial profile 
of the expected velocity dispersion \citep{Mam10}. So, we first find the mean velocity 
and estimate initial velocity dispersion using the $rms$ estimator. In successive steps we obtain stable 
and converging values of $\bar{\textrm{v}}$ and $\sigma_\textrm{v}$. This method yields a selection of 
104 cluster members, 3 and 45 foreground and background galaxies, respectively. Fig. \ref{fig:histogram} 
shows the redshift distribution of the galaxies in the range $0.06<z<0.11$ listed in table \ref{tab:catalog}.

The selection of 104 galaxy members shows a mean velocity $\bar{\textrm{v}}=22653 \pm 95$km s$^{-1}$ 
($z=0.0756$) and a $rms$ of $890 \pm 98$ km s$^{-1}$ (errors at 95\% c.l.) in the cluster rest frame.
In order to estimate a robust velocity dispersion, $\sigma_\textrm{v}$, we use the bi-weight scale 
estimator \citep{Bee90}, which is a procedure that offers satisfactory results for samples showing 
possible inhomogeneities. Applying this method to the 104 redshifts we obtain $845_{-90}^{+106}$ 
km s$^{-1}$. This result is in contrast with that obtained by \citet{Lop18}, who report a 
$\sigma_\textrm{v}=562_{-26}^{+31}$ km s$^{-1}$ using the bi-weight estimator. 
In our case, both $rms$ and bi-weight $\sigma_\textrm{v}$ estimations are in agreement within errors. 
However, in order to check the stability of $\sigma_\textrm{v}$ and discard possible deviations from the 
mean $\sigma_\textrm{v}$ along the cluster, we study the variation of this magnitude with the distance to 
the cluster center (assumed as \textit{BCG} position).

Fig. \ref{fig:3plots}, bottom panel, shows that the integral $\sigma_\textrm{v}$ profile is almost
completely flat for the whole cluster. This fact suggests that estimations of the $\sigma_\textrm{v}$ are
stable and robust even for radii as small as $r<0.2$ Mpc, which reveals that there are no obvious 
inhomogeneities in the velocity field \citep{Gir96}. This fact is also supported by the agreement between 
values obtained using the $rms$ and bi-weight estimators, as we see in the previous paragraph. However, in 
the following analyses we assume as more reliable value that $\sigma_\textrm{v}=845_{-90}^{+106}$ km 
s$^{-1}$, obtained using the bi-weight estimator given the robustness of this method in cases where the 
statistics clearly departs from the Gaussian distribution. The constancy of $\sigma_\textrm{v}$ 
variations in the velocity distribution, in particular it rules out a dependence of the mean velocity with 
the cluster-centric distance (see Fig. \ref{fig:3plots}, middle panel). We only notice a mild increase 
by about 650 km s$^{-1}$ for radius $r<0.25$ Mpc region. This fact may suggest that the dynamics 
of the cluster could be disturbed in its central region. We will analyze this fact in detail in following 
sections.

The BCG of RXCJ1111 (the ID 83) presents a velocity of 23428$\pm$5 km s$^{-1}$ (according to the SDSS
spectroscopic database), which is almost 800 km s$^{-1}$ higher with respect to the mean velocity of the 
cluster. \citet{Lop18} obtained a similar offset, 704 km s$^{-1}$. In addition to \textit{BCG}, which shows 
a magnitude $r'=14.17$, we identify two further bright galaxies more (IDs 60 and 67, with magnitudes $r'=14.55$ 
and 15.21), that we label as \textit{BCG2} and \textit{BCG3}, respectively. These two galaxies are located at 
1$^\prime$.30 and 2$^\prime$.25 toward the north and south of \textit{BCG}, respectively, configuring almost 
an alignment in the North-South direction. The X-ray surface brightness shows a double peak inside a 
SE-NW elongated inner region, and the \textit{BCG2} position coincides completely with the NW maximum of this 
double peaked emission (see Fig. \ref{fig:contours} and Fig. \ref{fig:BCG} left panel). On the other 
hand, the main BCG is shifted about 20 arcsec with respect to the SE X-ray peak, while \textit{BCG3} is 
placed to the south, where the X-ray profile shows an elongation in the external part. Therefore, 
the BCGs configuration is somehow linked to the X-ray diffuse emission of the cluster. This means that 
galaxies and the hot gas of the intra-cluster medium are interacting in some way. We analyse this 
interaction in sections \ref{sec:bcg-icm} and \ref{sec:mass}. 

We also detect 37 galaxies showing [OII] emission lines, labeled in table \ref{tab:catalog} as ELG 
(Emission Line Galaxy). The spectral resolution and SNR of our data allow us to detect [OII] emission 
lines with equivalent width >8 \AA. 10 out of 37 ELG galaxies are cluster members, 2 are placedkey
in the cluster foreground (with v$<20206$ km s$^{-1}$), while the rest (25 galaxies) are in 
the background (showing v$>25122$ km s$^{-1}$). So, the ELG members represent the 9.6\% of 
the cluster members in our sample. This is a typical fraction of ELGs in a cluster environment, 
which indicates that star-forming processes have been quenched in RXCJ1111, as expected
in high galaxy density environments and ICM showing high T$_X$ \citep{Lag08}.

\subsection{Velocity field}
\label{sec:velocity}

In general, any departure of the global velocity distribution along the LOS from a Gaussian is 
a reliable indicator that reveals the systems are dynamically disturbed or the presence of substructure 
(\citealt{Rib11}; \citealt{deCar17}). We measure skewness and kurtosis in order to investigate 
the shape of the velocity distribution of RXCJ1111. The skewness is related with the asymmetry 
of the velocity distribution, while the kurtosis indicates distributions presenting thinner/fatter 
tails. In our case, RXCJ1111 presents a velocity distribution which shows a skewness and kurtosis of 
$0.27 \pm 0.18$ and $-0.34 \pm 0.28$, respectively. Accordingly with our sample, 
we perform a Markov chain Monte Carlo (MCMC) method with 10000 simulations assuming a Gaussian profile 
with an average centre equal to zero and a standard deviation equal to one, sampled with 104 points. 
Errors were computed from the standard deviation of the values obtained. The positive skewness suggests 
that the velocity distribution is skewed to the right, while the kurtosis is almost 
compatible with zero. So, the global velocity distribution of RXCJ1111 is slightly asymmetric, with a skewness 
$1.5\sigma$ difference from zero. That is, the velocity distribution shows a small asymmetry relative to 
the normal Gaussian shape. This suggests that RXCJ1111 presents a dynamically disturbed state, 
probably dominated by two or more interacting substructures (see following sections). 

\begin{figure} 
\centering
\includegraphics[width=\columnwidth]{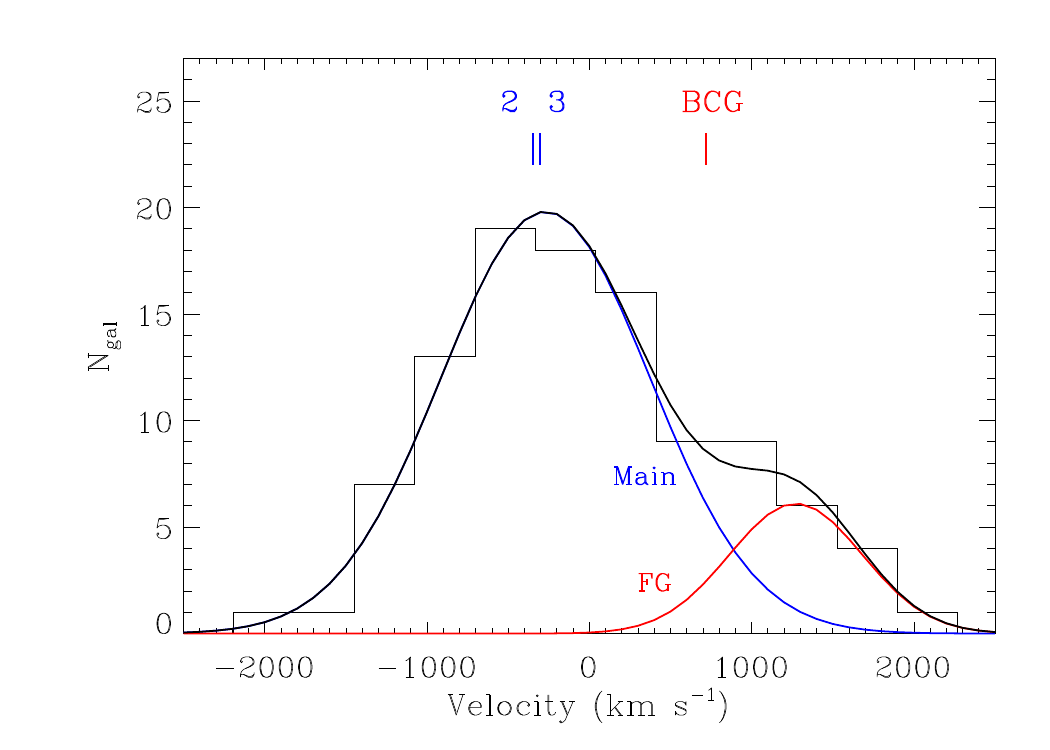}
\caption{The same distribution shown in the inner plot of Fig. \ref{fig:histogram} but
now the global fit (black curve) corresponds to two Gaussian components (in blue and red).
'\textit{BCG}' (in red), '2' and '3' (in blue) coloured labels agree with the most likely component.}
\label{fig:2gauss}
\end{figure}

The skewness obtained for the velocity distribution of RXCJ1111 support the
hypothesis that the cluster may be composed of two galaxy clumps, each of them showing its
Gaussian velocity distribution and both contributing to introduce distortions
in the global velocity distribution. With this idea in mind, we fit two Gaussian 
profiles to the global velocity distribution. The result is shown in Fig. \ref{fig:2gauss}.
Table \ref{tab:structures} lists the main parameters obtained for this two-component fit.
The best fit corresponds to two substructures with a difference in mean velocity of 
$\sim 1500$ km s$^{-1}$, one of them at $\sim -270$ km s$^{-1}$ with respect to the main velocity 
of RXCJ1111 with a $\sigma_\textrm{v}=644 \pm  56$km s$^{-1}$, which would constitute the
"main system". Additionally, we find a secondary substructure at $\sim 1270$ km s$^{-1}$ with respect to  
the cluster main velocity, with a $\sigma_\textrm{v}=410 \pm 123$ km s$^{-1}$, that we label as 
"FG"\footnote{We justify the name of this label in Sect. \ref{sec:FG}}. As it is shown in Fig. 
\ref{fig:2gauss}, \textit{BCG} presents a velocity offset of about $+1050$ km s$^{-1}$ with respect to the 
main system, while only $-550$ km s$^{-1}$ with respect to the secondary substructure. On the other 
hand, \textit{BCG2} and \textit{BCG3} are almost centred on the main system velocity distribution, showing 
only a velocity offset of about $-75$ km s$^{-1}$. Briefly, the velocity field of RXCJ1111 is 
consistent with a double cluster, where the secondary substructure contains a very bright 
galaxy, \textit{BCG}. On the other hand, \textit{BCG2} and \textit{BCG3} are part of the most populated 
component, the main body of the cluster. The membership relations between BCGs and the different 
substructures will be discussed in detail in Sect. \ref{sec:BCG_membership}.

\subsection{2D galaxy distribution}
\label{sec:spatial_distrib}

\begin{figure} 
\centering
\includegraphics[width=\columnwidth]{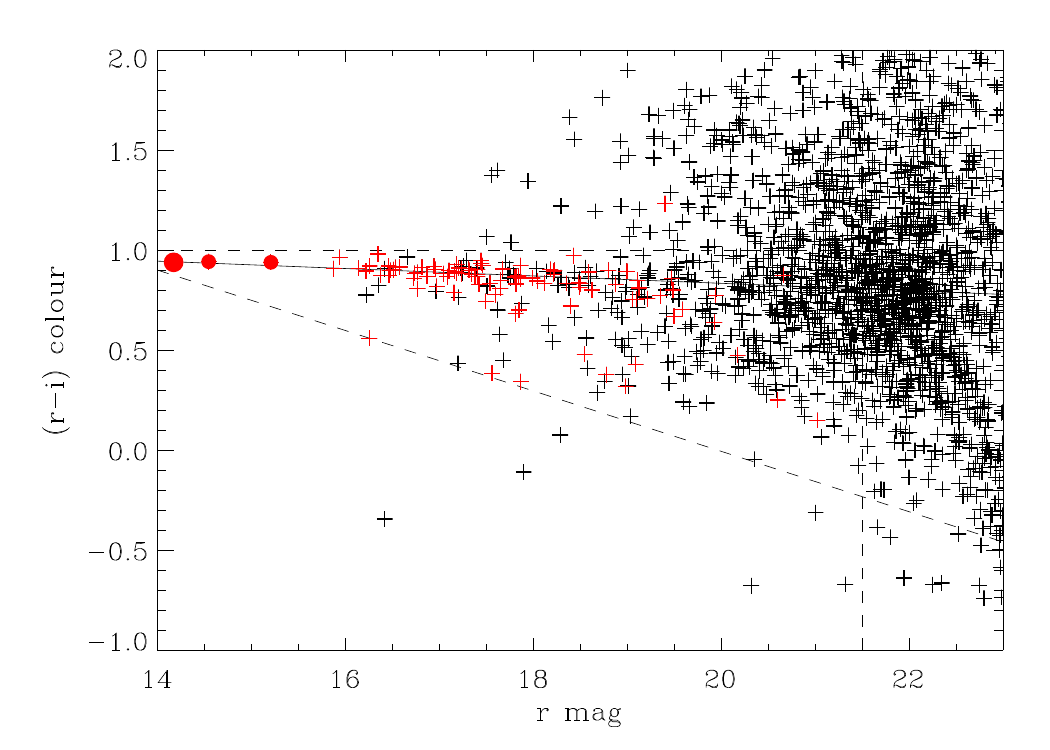}
\caption{Colour magnitude diagram ($g'-r'$, $r'$) of galaxies in a 
region of $12^\prime$ radius. Red symbols correspond to spectroscopically confirmed members. Large 
dots match with the three brightest galaxies, \textit{BCG}, \textit{BCG2} and \textit{BCG3}. Solid line represents the 
red sequence fitted $g'-r'=-0.185(\pm0.004) \times r'+1.20(\pm0.08)$. Dashed lines delimit the 
locus where likely members are selected.}
\label{fig:cmr}
\end{figure}

Spectroscopic samples suffer, in practice, from magnitude incompleteness. So, in order to get 
information of the galaxy distribution of the whole cluster, we adopt the photometric SDSS 
DR16 catalogues. Using the $g'$ and $r'$ $dered$ magnitudes, we construct the ($g'-r'$ vs $r'$)
colour-magnitude diagram (CMD) and we select likely members from the red sequence (RS) 
(see Fig. \ref{fig:cmr}) and "blue cloud" \citep{Gav10} following the technique detailed 
in \citet{Bar12}. The RS fitted follows the expression  $g'-r'=-0.185(\pm0.004) \times 
r'+1.20(\pm0.08)$ and we select both likely early-type and late-type galaxy members, residing 
in the RS and blue-cloud (below the RS), respectively. This locus is defined by $r'<21.5$ as 
magnitude completeness, the RS$\pm 3 \times rms$ as upper limit, and $-0.1509 \times r'+3.0125$ 
as lower limit in $g'-r'$ colour, respectively. This selection yields 926 likely members.

We use the likely members in order to explore the galaxy distribution. With this aim in mind, we
construct the contour levels of the isodensity galaxy distribution of the RXCJ111 likely member 
shown in Fig. \ref{fig:contours} (blue contours). This map has been obtained by computing the 
cumulative contribution of 926 small Gaussian profiles (with $\sigma=1$ arcsec width) positioned 
on each individual members over a grid of $258 \times 200$ points. The contour map obtained reveals
a double peak distribution clearly elongated in the NNE-SSW direction, which is oriented 
$\sim 25^{\circ}$ (see Sect. \ref{sec:spa_vel_correlat}) with respect to the North-South direction. 
The most significant peak is very close to \textit{BCG2}. It is important to note that both \textit{BCG2} and 
\textit{BCG3} galaxies are surrounded by many members and likely members, while \textit{BCG} is placed in a 
region where the galaxy distribution is not so high. We remark that RXCJ1111 presents a galaxy 
distribution and X-ray surface brightness profile almost coincident and following similar elongation 
and orientation. We will discuss this in detail in Sect. \ref{sec:optical_xray_mass}.

\subsection{Spatial-velocity correlations}
\label{sec:spa_vel_correlat}

In the past, many techniques have been developed to study the existence of substructures in clusters. 
One of the most successful procedures is the combined study of positions and velocities of galaxy 
members. The presence of different sub-clusters modifies the velocity field of galaxies, thus, by 
analyzing the space-velocity correlations, we can explore the internal kinematics of galaxy clusters.

In a first step, and given the evidence of bimodality exposed in Sect. \ref{sec:spatial_distrib}, 
we divide the galaxy members in two subsamples. The interaction between two substructures could induce 
inhomogeneities in the velocity distribution. Therefore, we look for significant gaps in the velocity 
histogram of RXCJ1111, which separates the galactic population of the two clumps. The most significant 
gap is detected around 600 km s$^{-1}$ (see Fig. \ref{fig:2gauss}). The existence of two 
substructures with v$<600$ km s$^{-1}$ and v$>600$ km s$^{-1}$ is supported by the fact that \textit{BCG2} 
and \textit{BCG3} would be associated to the low-velocity clump, while \textit{BCG} would be the brightest galaxy
of the high-velocity one. We analysed the spatial distribution of these two galaxy samples and found 
no evidence of spatial bimodality. This study suggests that the two galactic populations are 
intermingled at least in projection. However, the results found in Sect. \ref{sec:velocity} and the two Gaussian fit shown 
in Fig. \ref{fig:2gauss} strongly support the existence of two interacting substructures. 

\begin{figure} 
\centering
\includegraphics[width=\columnwidth,height=7cm]{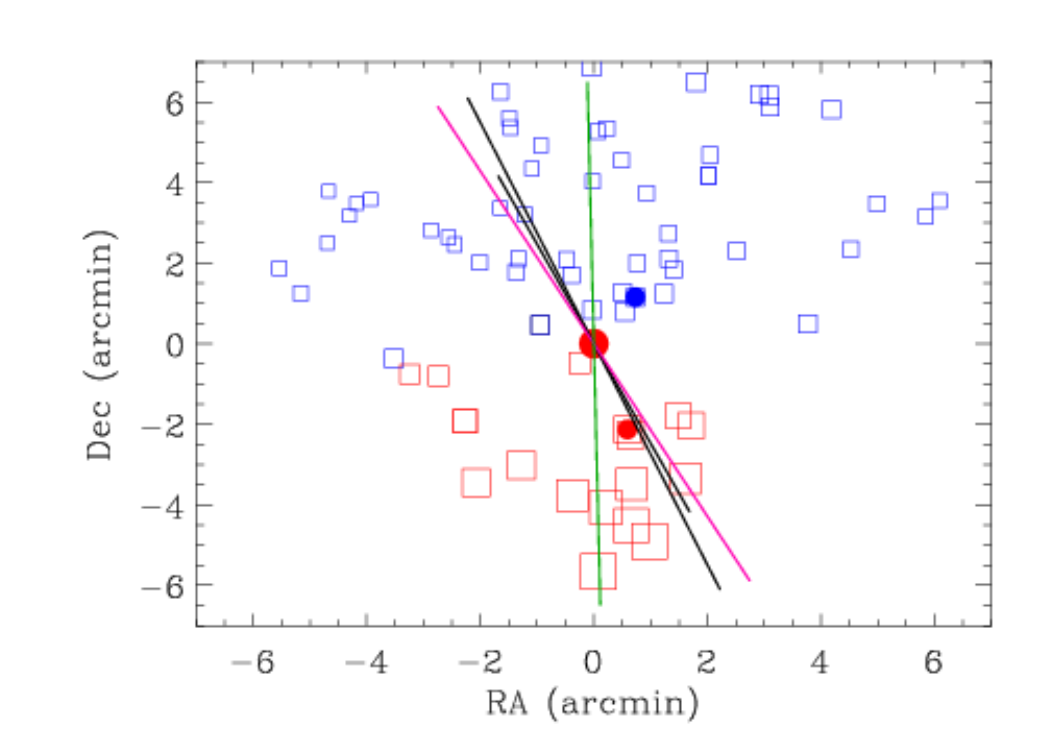}
\caption{Spatial distribution of the 78 cluster members inside a region of $7.5^\prime \times 7.5^\prime$, 
0.65 Mpc ($\sim 0.6r_{200}$) at the cluster redshift, from the cluster centre. Cluster member positions 
are marked with a square with size proportional to $exp(\delta_i)$ computed using the $\delta_i$ 
deviations obtained in the DS test. Red and blue correspond to galaxies with higher and lower 
$\delta_i$ deviations from the mean, $<\delta_i>$, respectively. The large and small black lines represent 
the directions of the velocity gradients of the whole sample in the inner region, respectively. Similarly,
magenta and green lines represent the orientations of the galaxy density distributions and X-ray surface
brightness map, respectively. \textit{BCG}, \textit{BCG2} and \textit{BCG3} positions are marked with filled dots.}
\label{fig:DS_delta}
\end{figure}

We perform a second test to check the existence of possible spatial-velocity segregation. We combine
galaxy positions and velocities by applying the classical $\delta$-statistics \citeauthor{Dress88} (DS)
test \citep{Dress88}, which identifies substructure searching for subsystems whose mean velocities 
and/or dispersion deviate from the global cluster values. After running this procedure using 1000
Monte Carlo simulations, we do not find any converging result. Fig. \ref{fig:DS_delta} shows the 
$\delta$-statistics test over a central region of $7.5^\prime \times 7.5^\prime$ size, which includes
78 cluster members. The outer regions are not considered in this test because they are not well enough 
sampled, and a lack of spatial sampling may introduce a biased result. We find a mean deviation of 
$0.22\pm0.18$, with a $p$-value statistics of 0.15, which is very low and means that there is not significant 
deviations. Thus, substructures are not spatially segregated significantly, and both galaxy populations 
are spatially mixed in the plane of the sky. However, note that galaxies in the south part of the cluster 
present higher $\delta_i$ respect the mean, while the northern area shows systematically lower $\delta_i$ 
deviations. Thus this test reveals an evidence of a clear velocity gradient in the North-South direction. 

In order to analyze the velocity gradient we fit a plane in the space-velocity frame. In a first step, 
we consider the full cluster member sample, obtaining the expression $\Delta\textrm{v}$=9.77x-26.91y+43.6,
where "x" and "y" are positions (following R.A. and Dec. coordinates, in arcmin respect to 
\textit{BCG}; positive values correspond to the north and west directions), which shows a gradient of 
325($\pm 350$) km s$^{-1}$ Mpc$^{-1}$ in a $110^\circ(\pm 8^\circ)$ angle (counterclockwise, from West to 
the North; $\Delta\textrm{v}$ takes positive values toward the south). In a second step, similarly, 
we only take into account the 78 cluster members inside the region of $7.5^\prime \times 7.5^\prime$ well 
sampled, and we find $\Delta\textrm{v}$=8.11x-20.54y+61.3, which presents a gradient of 254($\pm 465$) km 
s$^{-1}$ Mpc$^{-1}$ in a $112^{\circ}(\pm 14^\circ)$ angle. Therefore, velocity gradients are consistent in 
both cluster member samples, and this analysis confirms a slight increase of radial velocities 
toward the south part of the cluster and following the NNE-SSW direction. 

Fig. \ref{fig:DS_delta} also shows the orientations of the 2D spatial galaxy distribution of likely 
members and the X-ray surface brightness. We fit ellipses to the blue and white contours shown in Fig. 
\ref{fig:contours}, between $3.5^{\prime}$ and $7^{\prime}$ from \textit{BCG} to avoid the inner regions 
where we detect double peak profiles. This study reveals that galaxy distribution isocontours present 
a mean orientation\footnote{The West-East orientation corresponds to $0^{\circ}$ while $90^{\circ}$ 
point to North-South line.} of $115^{\circ} \pm 4^{\circ}$ (see magenta line in Fig. \ref{fig:DS_delta}), 
while X-ray contours are oriented $91^{\circ} \pm 6^{\circ}$ (green line in Fig. \ref{fig:DS_delta}). 
Therefore, the velocity gradient, the likely member distribution and the X-ray surface brightness 
are all oriented within $90^{\circ}-115^{\circ}$ angles, so the cluster shows a clear elongation in 
the NNE-SSW direction, produced by the overlapping galaxy populations showing slightly different velocity 
distributions. This finding support the fact that the cluster contains two galaxy clumps, one main 
body toward the north and a second substructure almost aligned in the line-of-sight, but slightly 
shifted toward the south. 

We try to associate individual galaxies to each substructure using 3D version of the Kaye's (KMM) 
Mixture Model algorithm \citep{Ash94}. This procedure separates the different components in velocity 
space, providing a probability that a given galaxy belongs to an individual component. The KMM 
algorithm needs a starting input configuration, so we provide two input lists. First, a list of 
galaxies associated with the southern cluster region. That is, the galaxies marked with red squares 
in Fig. \ref{fig:DS_delta}, which correspond to galaxies showing slightly higher $\delta_i$ in 
the DS-test. Second, a list of galaxies with v$>600$ km s$^{-1}$ with respect to the mean velocity of the
cluster. Our findings, running the KMM procedure on this two galaxy lists are not conclusive, 
and after running this algorithm iteratively we do not find a reliable result. The $p$-value 
to obtain this KMM result by chance is always higher than 0.23 (23\% probability). KMM 
procedure and DS-test are in agreement: none significant spatial segregation is detected for the 
main and secondary substructure. So, both galaxy populations are mixed in the plane of 
the sky. Thus, both populations seem to be almost completely aligned in the LOS.

\section{The BCG membership and substructure}
\label{sec:BCG_membership}

The three brightest galaxies of RXCJ1111 are also aligned in the North-South direction. However, 
the analysis of their radial velocities reveals that these three galaxies are not part
of a single mass halo. They follow very different kinematics. \textit{BCG2} and \textit{BCG3} 
are well centred on the velocity distribution of the main body of the cluster, while \textit{BCG} 
presents a velocity offset of about $990$ km s$^{-1}$. In contrast, \textit{BCG} shows 
a velocity offset of about $550$ km s$^{-1}$ with respect to the secondary substructure. So, from 
the dynamical point of view, \textit{BCG} seems to be linked to the secondary substructure. This 
evidence is supported by the KMM analysis performed in the previous section. In every KMM run, 
using different input configurations, we find that KMM always estimate a probability $>96\%$ 
that \textit{BCG} belongs to a secondary substructure, while for \textit{BCG2} and \textit{BCG3} we 
obtain a likelihood $>99\%$ for these two galaxies are part of main body of RXCJ1111. 

By studying a sample of 72 galaxy clusters showing SZ and X-ray emissions, L18 (see figure 6 
therein) found that only a negligible fraction ($<1$\%) of clusters may contain a BCG showing 
velocity offset as high as 1000 km s$^{-1}$ with respect to the main cluster velocity. Even for disturbed 
galaxy systems, this fraction is lower than $2$\%. In agreement with this result, \citet{Lau14}, on 
a sample of 178 clusters, also find that systems containing BCGs with peculiar velocities $>1000$ 
km s$^{-1}$ represent only a 2\% fraction (4 over 178), while the mean velocity offset of BCGs is about 
150 km s$^{-1}$ for clusters showing a velocity dispersion $\sigma_\textrm{v}\sim 600$ km s$^{-1}$. 
However, the velocity offset observed in the \textit{BCG} with respect to the secondary substructure of RXCJ1111,
is quite high, $550$ km s$^{-1}$, which suggests that this galaxy is greatly affected by other 
gravitational effects. One possibility is that \textit{BCG} could be suffering interaction with another 
mass halo, such as the \textit{BCG2} halo. In this way, \textit{BCG} and \textit{BCG2} could be orbiting each other, 
which is supported by the presence of a bar-like structure observed in X ray inner region.

To summarise, given the velocity offset observed in \textit{BCG} of RXCJ111, the most likely scenario would 
be that \textit{BCG} belongs to the secondary substructure. However, the substructure seems to be starting to 
interact with the main body of the cluster. On the other hand an interaction between  
\textit{BCG} and \textit{BCG2} haloes may explain the velocity offset observed in \textit{BCG} with respect its 
galaxy clump, the secondary substructure.

\subsection{The likely fossil substructure}
\label{sec:FG}

According to \citet{Jon03} and \citet{Dar10}, a galaxy system is considered fossil when it shows a 
magnitude gap between the most and the second brightest galaxy members greater than two, $\Delta 
m_{12} \ge 2$, within $0.5 \ r_{200}$. This magnitude gap arises naturally in undisturbed systems 
which have avoided infall into clusters, but where galaxy merging of the most luminous galaxies 
produces a extremly bright galaxy that dominates the core of the system. However, the precise value ($=2$) 
of the threshold in $\Delta m_{12}$ is quite arbitrary. The chance to find a value of $\Delta 
m_{12} >2$ in a typical Schechter function is very small. For instance, \citet{Zar14} checked this 
definition using spectroscopic information and only confirm five fossil groups showing $\Delta 
m_{12}>2$ in a sample of 34 systems previously identified as fossil systems by \citet{San07} using 
photometric samples. Anyway, despite the arbitrary definition of a fossil group, the observable 
$\Delta m_{12}$ is highly correlated with the evolutionary state of the system. This is confirmed 
by \citet{Zar21}. They find that radial orbits of galaxies as the cause of an increasing $\Delta 
m_{12}$ in groups. While relaxed systems show a large population of early-type galaxies with radial 
orbits (see \citealt{Biv03}; \citealt{Biv23}), except in the central regions, where dissipative 
friction may affect the dynamics of the brightest galaxies, clusters with smaller magnitude gap, 
showing more disturbed dynamical state and substructures, present more isotropic orbits. In other 
words, fossil systems are very relaxed structures, and this relaxation state is reflected in a 
large $\Delta m_{12}$.

In the case of RXCJ1111 and accordingly to the membership distribution
discussed in Sect. \ref{sec:BCG_membership}, \textit{BCG} is part of the secondary substructure, while 
\textit{BCG2} is the brightest galaxy of the main body of the cluster. Consequently, the $\Delta m_{12}$ of 
the main body is estimated as $\lvert r^\prime_{BCG2}-r^\prime_{BCG3} \rvert = 0.66$. On the 
other hand, in agreement with the velocity distribution, the fourth brightest galaxy, the ID 76, 
which shows a radial velocity of 21968$\pm$5 km s$^{-1}$, would be part of the main body, because
it shows a velocity difference of $\sim-1930$ km s$^{-1}$ with respect to the main velocity of the secondary
substructure. So, the $\Delta m_{12}$ of the secondary substructure could be only estimated with respect to 
the fifth brightest galaxy, the ID 103, which shows a similar velocity to \textit{BCG} one. In this way, 
$\Delta m_{12}$ can only be estimated as a minimum value, because there is a not negligible possibility
that the ID 103 galaxy is part of the main body of RXCJ111. Therefore, $\Delta m_{12}=\lvert 
r^\prime_{BCG}-r^\prime_{103} \rvert = 1.8$ in the secondary substructure.

From this analysis we can conclude that the secondary substructure is a fossil group, or at least 
an almost fossil system ($\Delta m_{12} =1.8 \sim 2$), which is now interacting with a more massive 
structure, the main body of the cluster. In Sect. \ref{sec:mass}, we will discuss the dynamics of 
this system. We will estimate the intervening dynamical masses and discuss a possible merging scenario.

\subsection{The interplay between BCGs and ICM}
\label{sec:bcg-icm}

\begin{figure} 
\centering
\includegraphics[width=\columnwidth]{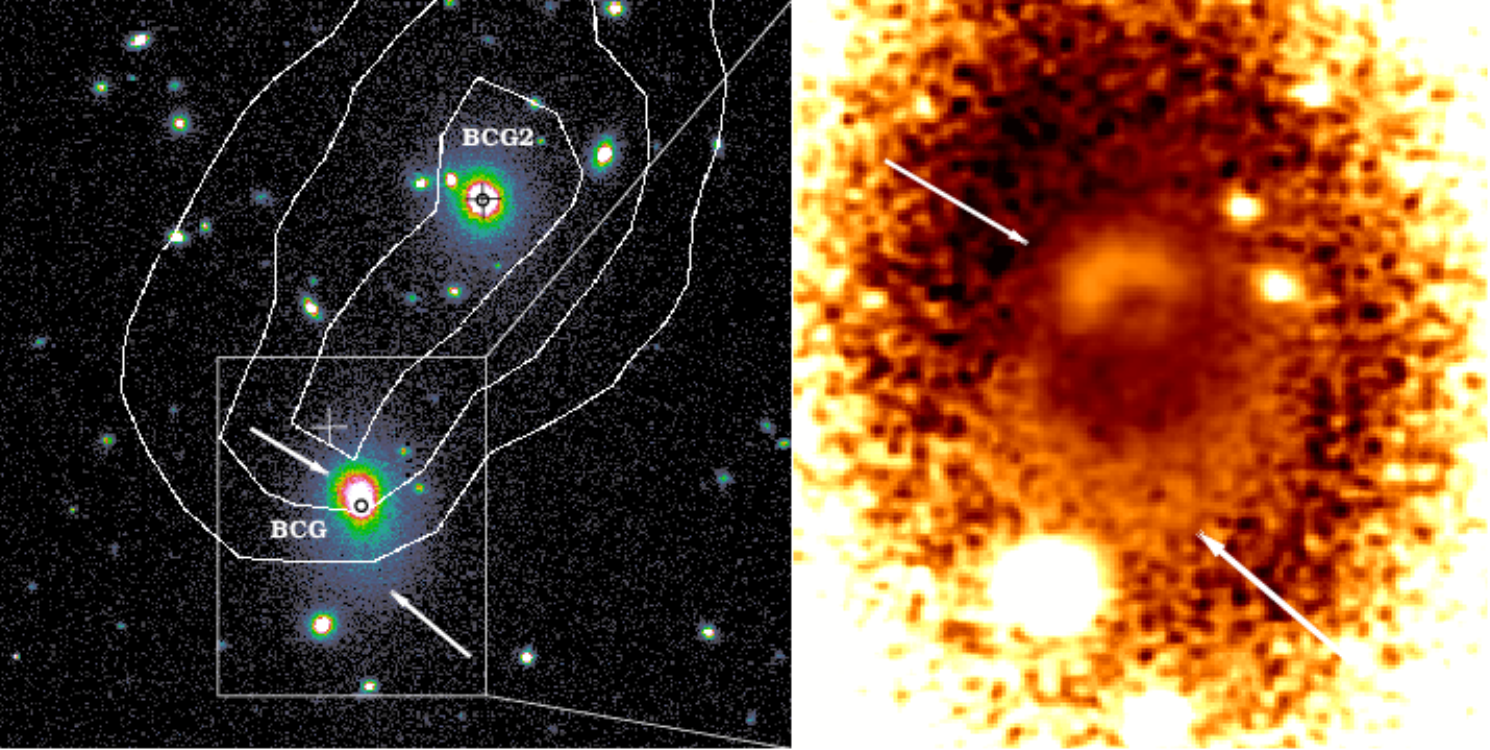}
\caption{Left: SDSS r-band image of the cluster core ($3^\prime \times 3^\prime$ field), 
around \textit{BCG} and \textit{BCG2}. White contours correspond to X-ray surface brightness. Black 
circles mark the galactic centres, while crosses indicate the X-ray peaks positions. Right: Residuals 
obtained after subtracting the best-fit model acquired with GASP2D to \textit{BCG} SDSS r-band 
image within a region of $1^\prime \times 1.4^\prime$. The arrows point the northern and southern 
shells surrounding the core of \textit{BCG}. Both images are oriented with north upward and 
east to the left.}
\label{fig:BCG}
\end{figure}

Cluster mergers are characterized by the presence of disturbed ICMs, but galaxies and ICM interact 
at different time-scales, each one revealing different dynamical properties. The hot gas component 
of the interacting systems often shows the presence of discontinuities in surface brightness and 
temperature that may not correlate with the most massive galaxies. Fig. \ref{fig:BCG}, left panel, 
shows the SDSS r-band image of the cluster core of RXCJ1111. Superimposed to this image are the 
innermost X-ray contours around \textit{BCG} and \textit{BCG2}. This inner contours show an 
elongated profile in the SE-NW direction, showing a double peak X-ray emission (marked with crosses 
in that figure), that connects \textit{BCG2} and \textit{BCG}. However, it is important to remark that, 
while the NW peak of the X-ray is the main one and perfectly coincide with \textit{BCG2} centre, the 
SE X-ray peak is substantially shifted with respect to \textit{BCG} position. This kind of missmatching 
between BCG positions and X-ray peaks is typically observed in cluster mergers (see i.e. \citealt{Lop18}). 
A clear and extreme case is seen most vividly in the case of the Bullet Cluster (\citealt{Bar02}, 
\citealt{Clo04}), where the gas has been separated from the galaxies during the core passage. A 
similar scenario can be taking place in RXCJ1111. The hot gas associated to the secondary structure, 
which is more concentrated around \textit{BCG}, has been shifted away, probably by ram pressure-stripping. 
In this way, the innermost hot gas of the secondary structure is starting to interact with the main 
body of RXCJ1111, producing a small displacement of the gas toward the outer regions of \textit{BCG}. 
We do not detect any evidence of the presence of shocks and cold fronts in the X-ray 
images, which could be an indication that the merger takes place close to the line-of-sight (see Sect. 
\ref{sec:optical_xray_mass}).

\subsubsection{The BCGs radio emission}
\label{sec:BCG_radio}

An interesting phenomenon observed in cluster merging processes is the presence of strong 
diffuse radio emission, forming extended halos and radio relics. So, with the aim of studying 
the radio emission in RXCJ1111, we inspect the public radio data archive. One of the few images 
available for this cluster is that obtained as part of the the VLA FIRST (Faint Images in the 
Radio Sky at Twenty-cm) survey \citep{Bec95} in the frequency 1.4 GHz, with a exposure time of 
180 sec and using a beam size of 5.4 arcsec.

The contour levels of the VLA radio image are shown in Fig. \ref{fig:contours}, upper right 
corner, which only reveals radio emission from the three brightest galaxies. The same three
radio sources were reported in \citet{Owe93} and \citet{OL97}, where no trace of diffuse 
emission was detected from a cluster halo. The emission observed corresponds to classical 
radio galaxies with lobes created by jets filled with relativistic plasma \citep{Beg84}. \textit{BCG} 
presents a very weak central emission, while that from \textit{BCG2} and \textit{BCG3} is more extended. The 
emission corresponding to the \textit{BCG2} resembles a classical head and tail source. That is, we 
see central emission and lobes dragged to the south-west. On the other hand, the \textit{BCG3} presents 
lobes that are still on both sides of the galaxy, but are not strictly aligned with the galaxy 
centre. This fact suggests that \textit{BCG3} is moving with respect to its surrounding medium.

The fact that no diffuse and extended emission has been detected implies that the energy 
contribution of potential relativistic electrons in the ICM is not large enough to be detected.
In consequence, we may infer that the merging process is still in a very early phase. However, more 
deep radio observations would be necessary in order to completely confirm this hypothesis.

\begin{table*}[!htbp]
\caption{Global properties for the whole cluster and clump components detected in RXCJ1230.}
\begin{center}
\begin{threeparttable}[t]
\begin{tabular}{lcccccc}
\midrule \midrule
Structure  & $\bar{\textrm{v}}$ & $\sigma_\textrm{v}$ & N$_{gal}$ & M$_{200}$			  & M$_{500}$                     & r$_{200}$	       \cr
           & (km s$^{-1}$)      & (km s$^{-1}$)       &           & ($\cdot10^{14}$ M$_{\odot}$) & ($\cdot10^{14}$ M$_{\odot}$) & ($h_{70}^{-1}$Mpc) \cr
\midrule
Global     & $22653\pm 95$ & $845_{-90}^{+106}$ &       104 & $3.3 \pm 1.0$ & $2.1 \pm 0.7$ & $\sim 1.4$ \cr
Main       & $22364\pm 54$ & $644 \pm  56$      & $\sim 78$ & $1.9 \pm 0.4$ & $1.2 \pm 0.6$ & $\sim 1.2$ \cr
FG         & $24023\pm134$ & $410 \pm 123$      & $\sim 26$ & $0.6 \pm 0.4$ & $0.3 \pm 0.2$ & $\sim 0.8$ \cr
\midrule
\end{tabular}
\begin{tablenotes}
\item \footnotesize{Note: N$_{gal}$ of the main system and FG subcluster have to be
taken as guide values. Galaxies with $\bar{\textrm{v}}<600$ and $\bar{\textrm{v}}>600$ km s$^{-1}$ are assumed to belong 
to the main clump and FG substructure, respectively. The masses are dynamical ones.
}
\end{tablenotes}
\end{threeparttable}
\end{center}
\label{tab:structures}
\end{table*}%

\subsubsection{\textit{BCG} photometric structure}

We performed a two-dimensional photometric decomposition of \textit{BCG} using the GAlaxy Surface 
Photometry 2 Dimensional algorithm \citep[GASP2D,][]{mendezabreu08,mendezabreu14}. To this aim, 
we used the SDSS r-band image. We modelled the surface brightness distribution of the galaxy 
assuming both a single S\'ersic \citep{sersic68} and a S\'ersic+Exponential distributions. GASP2D 
returns the best-fitting values of the structural parameters of each morphological component by 
minimizing the $\chi^2$ after weighting the surface brightness of the image pixels according to 
the variance of the total observed photon counts due to the contribution of both galaxy and sky 
\citep[see also][]{mendezabreu17}. We obtained a best fit in terms of the $\chi^2$ when using the 
S\'ersic+Exponential model. In principle, this would allow us to fit the low surface-brightness 
and extended component that BCGs could present \citep[e.g.,][]{nelson02,mendezabreu12}. However, 
according to the Bayesian Information Criterion \citep[BIC,][]{schwarz78} we found that 
adding an extra component to the S\'ersic fit did not statistically improve our results.

The right panel of Fig. \ref{fig:BCG} shows the residuals after subtracting our best S\'ersic 
model to the original SDSS r-band image. One can clearly see the presence of two non-axysimmetric 
structures. The first one is surrounding the galaxy core and the second one, more extended, resemble 
the presence of a tidal tail towards the southern region of the galaxy. It is worth noticing that 
both non-axysimmetric structures likely share the same north-south direction and they were also 
present when modelling the surface brightness with a S\'ersic+Exponential model.

We argue that the residuals observed in \textit{BCG} of RXCJ1111 are likely due to the merger with one 
or two galaxies. The presence of double nucleii due to the merger of galaxies is common fact that
can be studied using photometric \citep{komossa03,benitez13} and spectroscopic \citep{patton16} 
methods. Our model generated by GASP2D does not reveal signs of a double nucleus in \textit{BCG} core. 
However, the excess of light (over the S\'ersic model) right to the south of the galaxy center 
resembles what would be the final stages of a merger with a relatively massive and concentrated 
companion \citep{hendel15}. The outer non-axysimmetric structure has a different apparent shape 
in its northern and southern part with respect to the galaxy center. The former resembles the 
typical shells formed during a minor merger with a relatively gas-poor companion \citep{mancillas19}. 
The southern part shows a tidal tail shape, which could also have been created by a minor merger, but on 
a less radial orbit with respect to the shells. In summary, even if the details of the past mergers 
suffered by \textit{BCG} are difficult to unveil using only the available images, it is clear that 
\textit{BCG} shows several signs of recent and past interactions with other galaxies. The structures 
described in this section cannot be produced due to interactions with the ICM, fast encounters 
with other cluster galaxies (harassment), nor are the result of a merger with another bright galaxy. 
Therefore, within the context described in this paper of the \textit{BCG} being the central galaxy of a 
transitional FG, we suggest that it is the result of several galaxy mergers that might be 
leading to observed the system as a FG.

\section{X-ray properties}
\label{sec:Xprop}

We extracted X-ray surface brightness profiles after the subtraction of the X-ray sky and 
instrumental background. For RXCJ1111 we determined one surface brightness profile for the entire 
cluster assuming approximate spherical symmetry. We fitted the profiles with $\beta$- and double 
$\beta$-models. We found that single $\beta$-models provide a good fit to the outer 
parts of the profiles, from which we determined the gas density and gas mass profiles. 
Only for the southern part of RXCJ1111 the double $\beta$-model provides an interesting alternative.
The fit parameters for the profiles are given in Table \ref{table:r1111}. M$_{500}$ is the 
mass inside $r_{500}$, M$_{gas}$ is the gas mass inside the same radius. We do not detect 
any evidence of the presence of shocks and cold fronts in the X-ray images, which could be 
taken as an indication of an advanced merger stage. But we show below that the merger occurs probably 
close to the line-of-sight and in this case it is extremely difficult to see the signatures of shocks 
and to draw conclusions from the present data. $f_{gas}$ is the gas mass fraction of the total mass, 
$r_{500}$ is the radius projected in the sky, L$_X$ is total X-ray luminosity, M$_{500}($L$_X)$ is 
the mass estimated from the L$_X - $M relation, T$_X$ the measured X-ray temperature for the cluster 
region inside $r_{500}$, $r_c$ the core radius, and $\beta$ the slope parameter of the electron 
density profile.

\begin{figure}
\centering
\includegraphics[width=\columnwidth]{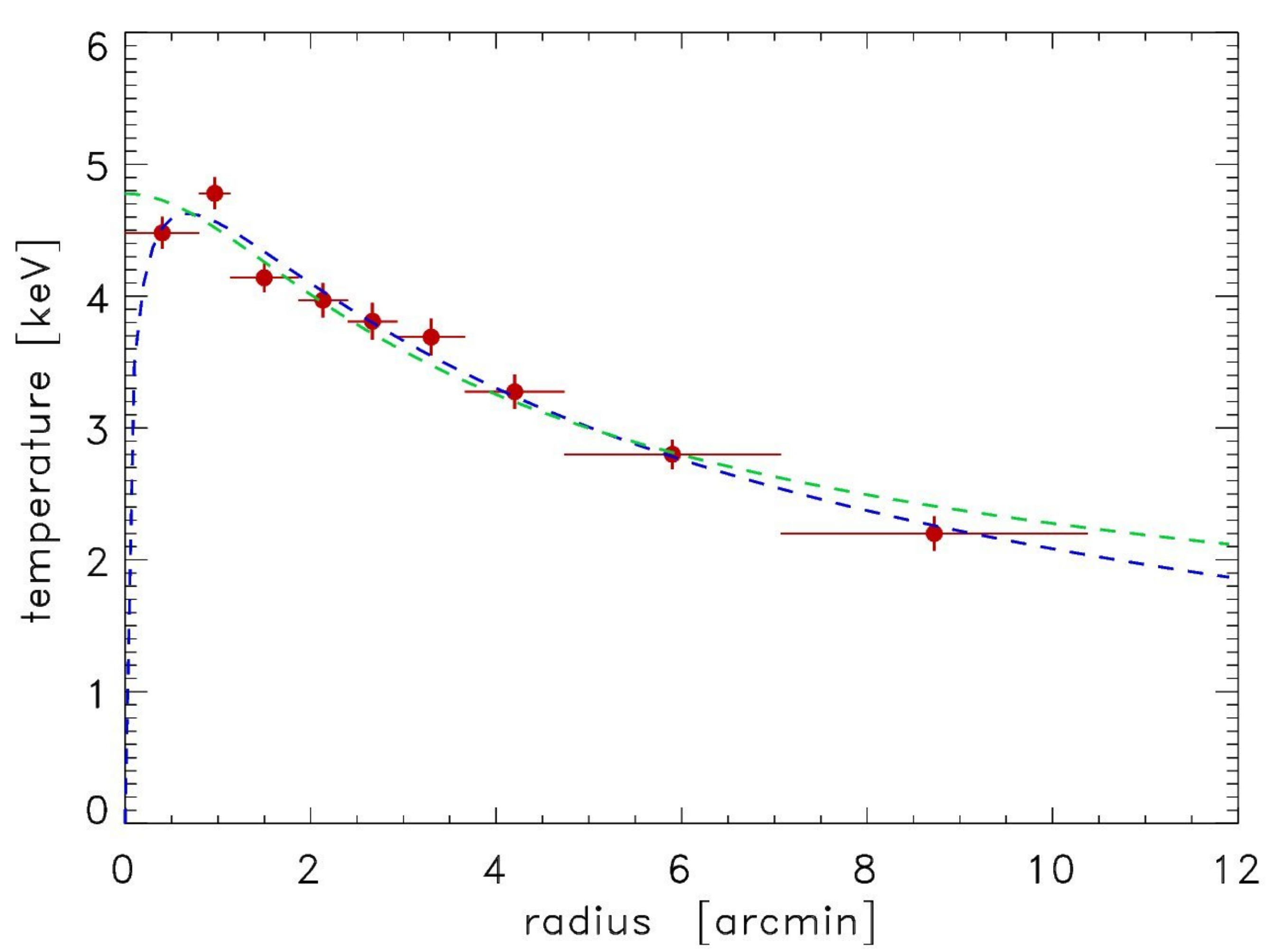}
\caption{Observed temperature profile of RXCJ1111 fitted by Eq. 1 (blue curve) and by a polytropic model (green curve).}
\label{fig:r1111_tprof}
\end{figure}

The temperature of the ICM was also determined from the analysis of the background-subtracted 
X-ray spectrum. We determined the temperature in nine concentric radial bins for the cluster 
RXCJ1111. The bins were constructed with a minimum number of 2000 photons per bin. The resulting 
temperature profile is shown in Fig \ref{fig:r1111_tprof}. The sudden temperature drop of 
the fitted profile at the centre is an artifact caused by the peculiar location of the innermost 
data points and is certainly not real and ignored in the following analysis. The outer temperature 
profile of RXCJ1111 can be approximated by a polytropic model with a $\gamma$ parameter of $\sim 1.2$.
We also use the fitting formula of \citet{Vik06} to approximate the temperature profile.
In this formula we drop the part which describes the central temperature decrease which is an effect 
of a cool-core, which is not relevant for RXCJ1111. The equation applied has thus the form.

\begin{equation}
\label{eq:T}
    T(r) = T_0~\frac{(r/r_c)^{-a}} {(1 + (r/r_c)^b)^{c/b}}
\end{equation}

In addition, we determined temperatures from spectra extracted over the entire region inside 
$r_{500}$ and $0.75 r_{500}$ with results provided in Table \ref{table:spec_prop}.
In Col. 1, A and B regions correspond to $r<r_{500}$ and $r<0.75r_{500}$ circles, 
respectively, while A$_{\rm X}$ and B$_{\rm X}$ are annuli where the central region (Core), 
inside r=0.15$r_{500}$, was cut out. "Z" is the Fe abundance (in solar units with 
abundances from \citealt{asplund09}), "Norm" is the APEC (Astrophysical Plasma Emission Code; 
\citealt{Smi01}) normalisation, and f$_X$ and L$_X$ are the flux and luminosity in the [0.5--2.0] 
keV band, respectively. 

\begin{table}[ht]
\caption{X-ray properties of RXCJ1111}             
\label{table:r1111}
\centering          
\begin{tabular}{lcc}     
\hline
\hline
 & {\rm Single beta model} \\
\hline
M$_{500}$ $(\times 10^{14}$ M$_{\odot}$)      & $1.68 \pm0.25$ \\
M$_{gas}$ $(\times 10^{13}$ M$_{\odot}$)      &  1.97          \\
f$_{gas}$                                     & 11.3\%         \\
r$_{500}$ (arcmin)                            &  9.07          \\
L$_X$ ($\times 10^{44}$ erg s$^{-1}$)         & 0.5            \\
M$_{500}(L_X)$ $(\times 10^{14}$ M$_{\odot}$) & 2.5            \\ 
T$_X$ (keV)                                   & $\sim 3.6$     \\
$n_{e0}$ (cm$^{-3}$)                            & 3.6$\cdot 10^{-3}$\\
r$_c$                                         & 1.77           \\
$\beta$                                       & 0.673          \\
\hline                 
\end{tabular}
\end{table}

\begin{table}[ht]
\caption{X-ray spectral properties of RXCJ1111}
\label{table:spec_prop}
\centering
\begin{tabular}{llllll}
\hline
\hline
Reg. & T$_X$   & Z & Norm & f$_{\rm X}$ ($\cdot 10^{-12}$)     & L$_{\rm X}$ ($\cdot 10^{43}$)\\ 
     & (keV)   &   & $\cdot 10^{-3}$)     & (erg/s/cm$^2$) & (erg/s) \\
\hline
A            & 3.635 & 0.291 & 6.43 & 3.3   & 5.02 \\
A$_{\rm X}$  & 3.323 & 0.227 & 4.59 & 2.26  & 3.45 \\
B            & 3.74  & 0.305 & 6.25 & 3.16  & 4.81 \\
B$_{\rm X}$  & 3.360 & 0.224 & 4.05 & 1.99  & 3.03 \\
Core         & 4.489 & 0.443 & 2.14 & 1.11  & 1.68 \\
\hline
\end{tabular}
\end{table}

\begin{figure}
\centering
\includegraphics[width=\columnwidth]{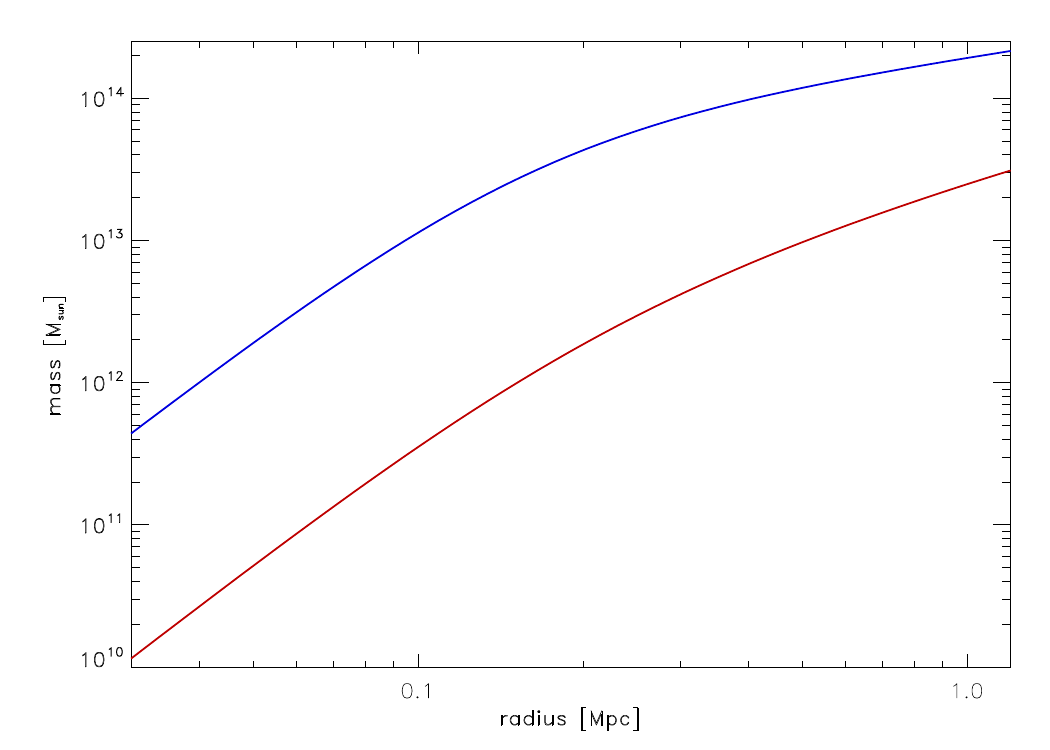}
\caption{Gravitational (blue) and gas mass (red) profile of RXCJ1111.}
\label{fig:r1111_mprof2}
\end{figure}

Based on the density and temperature profile, we determined the gas mass and total mass 
profile, assuming hydrostatic equilibrium for the derivation of the latter. We find a cluster 
mass inside $r_{500}$ for RXCJ1111 of M$_{500} = 1.68 \pm0.25 \times 10^{14}$ M$_{\odot}$. 
The results based on the temperature profile given by Eq. \ref{eq:T} and by the polytropic model 
agree within the error bars. Mass and gas mass profiles determined from the best fit for 
RXCJ1111 are shown in Fig. \ref{fig:r1111_mprof2}. For the X-ray luminosity of the cluster 
in the 0.5 to 2 keV energy band, we obtain a value of $L_{X,500} = 5 \times 10^{43}$ erg 
s$^{-1}$. Based on the mass-luminosity relation proposed by \citet{Pra09}, we 
expect for this value a cluster mass of about M$_{500} = 2.5 \times 10^{14}$ M$_{\odot}$. 
For the mean temperature derived for the entire cluster inside $r_{500}$ of 3.6 keV we would 
expect a cluster mass of about M$_{500} = 2.4 \cdot 10^{14}$ M$_{\odot}$ (e.g. M-T relation 
of \citealt{Arn05}). The mass implied by the temperature and X-ray luminosity of the 
intra-cluster medium for the case of a more relaxed cluster is therefore higher than the mass 
determined from the detailed X-ray analysis. This discrepancy can, for example, be explained 
by the merger state of the cluster.

\section{Dynamics of RXCJ1111}
\label{sec:mass}

As it has been argued in previous sections, we can conclude that RXCJ1111 is formed by one main 
body and one likely fossil substructure, almost completely aligned in the LOS. Their galactic 
population are superimposed in the plane of the sky, but from the velocity analysis we can affirm 
that the brightest galaxy, \textit{BCG}, can only be part of the secondary substructure, while \textit{BCG2} 
and \textit{BCG3} belong the main body. The X-ray surface brightness map suggests that they are starting to 
interact. In this section we estimate the dynamical masses and virial radii in order to 
characterize this complex.

RXCJ1111 shows a velocity distribution (see Fig. \ref{fig:2gauss}) that allows us to distinguish
between the two galaxy clumps. On the other hand, we detect a $\sim$30\% discrepancy in the mass
estimates derived from the temperature profile and the M-T relation (see previous section), which could 
be due to a X-ray temperature increase. Taking into account these two facts, is reasonable to assume 
that the fossil substructure has not yet completely merged into the cluster and both galaxy haloes still 
keep a roughly dynamical equilibrium. That is, velocity distribution of galaxies
are not so disturbed, but the gas temperature starts to increase. In this scenario, it is possible to estimate 
virial dynamical masses and radii, and compute the mass of the whole cluster as the sum of the 
individual masses. Table \ref{tab:structures} lists the main properties of RXCJ1111 and summarises 
the main velocities and dispersions detailed in Sect. \ref{sec:velocity}. 

Galaxies are embedded in the gravitational potential of the cluster, thus their velocities can be 
used to estimate the dynamical mass of galaxy clumps. In this way, we use the velocity dispersion,
$\sigma_\textrm{v}$, and its relation with the virial mass, M$_{200}$, to estimate the dynamical 
mass of the components of RXCJ1111. One of the most common ways to determine dynamical masses of 
clusters from their velocity dispersion is using scaling relations. In the literature, there are 
many examples offering $\sigma_\textrm{v} - $M$_{200}$ relations\footnote{Notice that dynamical 
masses derived from velocity dispersion always present high errors due to the fact that $\sigma_v$ 
is cubed in the M$_{200}-\sigma_v$ relation.} (see e.g. \citealt{Evr08}; \citealt{Saro13}; \citealt{Mun13}; 
\citealt{Ferra20}). All of them provide similar values, however 
we follow the \citet{Ferra20} relation. This procedure is calibrated using the \cite{Mun13} simulations, 
which consider not only dark matter particles but also subhalos, galaxies and AGN feedback. 
\citet{Ferra20} go one step further and consider statistical and physical effects in samples 
containing small numbers of cluster members. Following this prescription, we find 
dynamical masses of M$_{200}=1.9 \pm 0.4 \times 10^{14}$ M$_{\odot}$ and $0.6 \pm 0.4 \times 10^{14}$ 
M$_{\odot}$ for the main cluster and the secondary substructure, respectively. In order to compare 
these values with that obtained using X-ray emission, we convert M$_{200}$ into M$_{500}$ following 
the relation given by \citet{Duf08}. That is, rescaling M$_{500}$ from M$_{200}$ assuming a 
concentration parameter $c_{200}=4$ (a suitable value for clusters at $z<0.1$ and M$_{200} \sim 
10^{14}$ M$_{\odot}$), integrating a Navarro-Frenk-White (NFW) profile \citep{NFW97} and 
interpolating to obtain M$_{500}$. So, we obtain M$_{500}=1.2 \pm 0.6 \times 10^{14}$ M$_{\odot}$ 
and $0.3 \pm 0.2 \times 10^{14}$ M$_{\odot}$ for the main body of the cluster and the substructure, 
respectively. Therefore, taking into account these values, we can conclude that RXCJ1111 presents 
a total dynamical mass of M$_{200}=2.5 \pm 0.6 \times 10^{14}$ M$_{\odot}$ and M$_{500}=1.5 \pm 0.6 
\times 10^{14}$ M$_{\odot}$, estimated as the sum of the individual masses of the two clump components. 

Quasi-virialised regions can be determined by evaluating the virial radius of each galaxy clump.
This radius is usually estimated as the radius of a sphere of mass M$_{200}$ inside which the 
matter density is 200 times the critical density of the Universe at the redshift of the system, 
$200\rho_c(\textrm{z})$. Therefore, M$_{200}=100$r$_{200}^3 \textrm{H}(\textrm{z})^2/$G. So, 
following this expression, we obtain r$_{200} \sim 1.2$ and 0.8 $h_{70}^{-1}$ Mpc, for the main 
body and the substructure, respectively. We compile virial masses and radii in Table \ref{tab:structures}.

As an exercise, we could compare the total dynamical mass of RXCJ1111 above obtained with that 
determined assuming the cluster as a single galaxy clump, neglecting its substructured composition.
In this way, in Sect. \ref{sec:members_global} we saw that RXCJ1111 presents a global velocity 
dispersion $\sigma_\textrm{v}=845_{-90}^{+106}$. Applying the $\sigma_\textrm{v} - $M$_{200}$ 
relation proposed by \citet{Ferra20}, we obtain a global dynamical mass M$_{200}=3.3 \pm 1.0 \times 10^{14}$ 
M$_{\odot}$ and M$_{500}=2.1 \pm 0.7 \times 10^{14}$M$_{\odot}$. Comparing these values with that above 
obtained assuming substructure, we see that both numbers agree within $1\sigma$ error. However,
the results derived from a global velocity dispersion seems to be slightly higher. This fact may be an 
indication that dynamical masses derived from a global velocity dispersion is overestimated for
clusters whith evident substructure.

\subsection{Explaining the optical and X-ray properties}
\label{sec:optical_xray_mass}

By analysing X-ray data (see Sect. \ref{sec:x-ray}), we find an elongated configuration. The main
X-ray emission comes from the north part of the cluster, while a bright extension to the south
is also observed. Therefore, on the one hand, X-ray analyses confirm that RXCJ1111 presents an
unrelaxed state, and on the other, the mean X-ray emission matches with the north part of the 
cluster, which includes a emission maximum centred on \textit{BCG2}. However, the fact that we do 
not see a strong X-ray peak in the centre and rather a smeared peak could also be due to the 
orbiting central galaxies with their DM halos. In the optical, this scenario would explain the
large offset ($\sim -550$ km s$^{-1}$) of \textit{BCG} respect the main velocity distribution of
its corresponding clump, the secondary substructure. Two mechanisms can help to retard 
the \textit{BCG}. The first is dynamical friction (Merritt \citeyear{merr83}; \citeyear{merr84}; 
\citeyear{merr85}), which is generally strong for a massive BCG, but has most impact when the 
galaxy is near the center of the primary cluster. The second mechanism is simply deflection of 
the BCG in its orbit as the result of a nonzero impact parameter. In the latter case, there is 
no loss of orbital energy, but the component of the BCG's velocity projected onto the line of 
sight is reduced. Neither effect (dynamical friction or deflection) can be significant early in 
the merger. However, the velocity offset of \textit{BCG} provides a strong argument that 
this merger has advanced at least close to core passage.

Once the point-like sources are removed, the rest of the X-ray diffuse emission on small scales is 
consistent with statistical fluctuations, which is difficult to observe if the merger 
happens in a direction close to the line-of-sight. However, more interesting is the 
agreement between X-ray shape emission from hot gas and galaxy spatial distributions. 
Both galaxy distribution and X-ray morphology suggest that the merging is happening almost along 
the LOS, maybe with a small impact parameter to the south (see Sect. \ref{sec:merging}).

We find that M$_{500}$ derived from mass-Xray luminosity relation and from T$_X$, M$_{500}(L_X)$ and 
M$_{500}(T_X)$ are about $\sim 2.5$ and $\sim 2.4 \times 10^{14}$ M$_{\odot}$, respectively. These values are 
slightly higher respect M$_{500,X}$ and M$_{500,dyn}$. Probably, the reason behind this 
discrepancy is that RXCJ1111 is in a early phase of a merging (see Sect. \ref{sec:merging}). 
Thus, an on-going merging may be producing a small increase in the temperature of the hot gas 
of the ICM. Similarly, the L$_X$ is also enhanced due to effect produced by the collision of 
substructures and gas compression.

One of the most interesting question arises from the comparison between masses derived 
from X-ray and galaxy dynamics. Assuming two separate galaxy clumps, we obtain M$_{500,dyn}=1.5 \pm 
0.6 \times 10^{14}$ M$_{\odot}$ and M$_{500,X}=1.68 \pm 0.25 \times 10^{14}$ M$_{\odot}$ from X-ray 
and galaxy dynamics, respectively, which are in quite good agreement within $1\sigma -$ errors. 
However, assuming RXCJ1111 to be composed by only one galaxy population, we obtain M$_{500,dyn}=2.1 
\pm 0.7 \times 10^{14}$ M$_{\odot}$ that seems to be slightly higher, but within $1\sigma$, 
respect to M$_{500,X}$. This fact confirms that the most appropriate way to estimate realistic 
dynamical masses, even in not so relaxed clusters, is to identify the cluster components and consider 
each galaxy clump separately, so estimating masses within regions showing a more relaxed state. 
This method validates similar techniques to determine dynamical masses in a more accurate way, 
which have been successfully applied for merging clusters in the past (e.g. \citealt{Gir08}; 
\citealt{Bos12}).  

Summarizing, the mass estimates derived from global dynamics of galaxies, the mass from global T$_X$ 
and L$_X$ are somehow overestimated, probably due to the unrelaxed state of RXCJ1111. Hence the 
importance of estimating dynamical masses in non-virialised clusters taking into account galaxy 
clumps separately. In addition, a good knowledge of the dynamical state is crucial in order to 
derive realistic X-ray properties in this sort of clusters.

\subsection{A 3D merging model}
\label{sec:merging}

As we pointed out above, both the main cluster and the fossil substructure can be separated
in the velocity field, but not in the spatial distribution. The on-going collision will involve 
two mass halos with a mass ratio of 3:1. On the other hand, typical X-ray temperatures of 
relaxed clusters with M$_{500} \sim 1.5 \times 10^{14}$ M$_{\odot}$ are about 2.1 keV (see 
e.g. Fig.9 in \citealt{ket13}) and references therein), however, we measure a global T$_X\sim 3.6$ 
keV. Thus, we measure a $\sim \Delta$T$_X\sim 1.5$ keV enhancement in the X-ray temperature of the
ICM that may also be explained by the fact that the main cluster and the fossil substructure 
are starting to collide. Galaxy clusters showing mergings in advanced states present an 
ICM much more disturbed with very high X-ray temperature (see e.g. \citealt{Bos06}; 
\citealt{Bar02}). In the following we propose a merging model in order to explain the 3D 
dynamics of this two-body collision.

The relative dynamics of RXCJ1111 is relatively simple and basically described by a interacting 
main body and substructure with mass ratio about 3:1. We analyze this interaction from different
approaches, based in an energy integral formalism and considering a flat space-time and Newtonian 
gravity \citep[see e.g.][]{Bee82}. The three most important observables in a two-body interaction
are: the total mass of the two systems computed as the sum of the individual components, M$_{200,sys} 
\sim 2.5 \pm 0.6 \times 10^{14}$ M$_{\odot}$ (see Sect. \ref{sec:mass}); the relative LOS 
velocity in the cluster rest frame, $V_r=1540 \pm 135$ km s$^{-1}$; and the projected physical 
distance. This last term is quite undefined because both galaxy clumps seem to be superimposed 
in the plane of the sky. However, we may assume a projected distance of 82 arcsec $=0.12$ 
$h_{70}^{-1}$ Mpc, which is the separation between the two intervening brightest galaxies,
\textit{BCG2} and \textit{BCG}, assumed to be the gravitational centres of the main cluster and the 
substructure, respectively.

\begin{figure} 
\centering
\includegraphics[width=9cm,height=6cm]{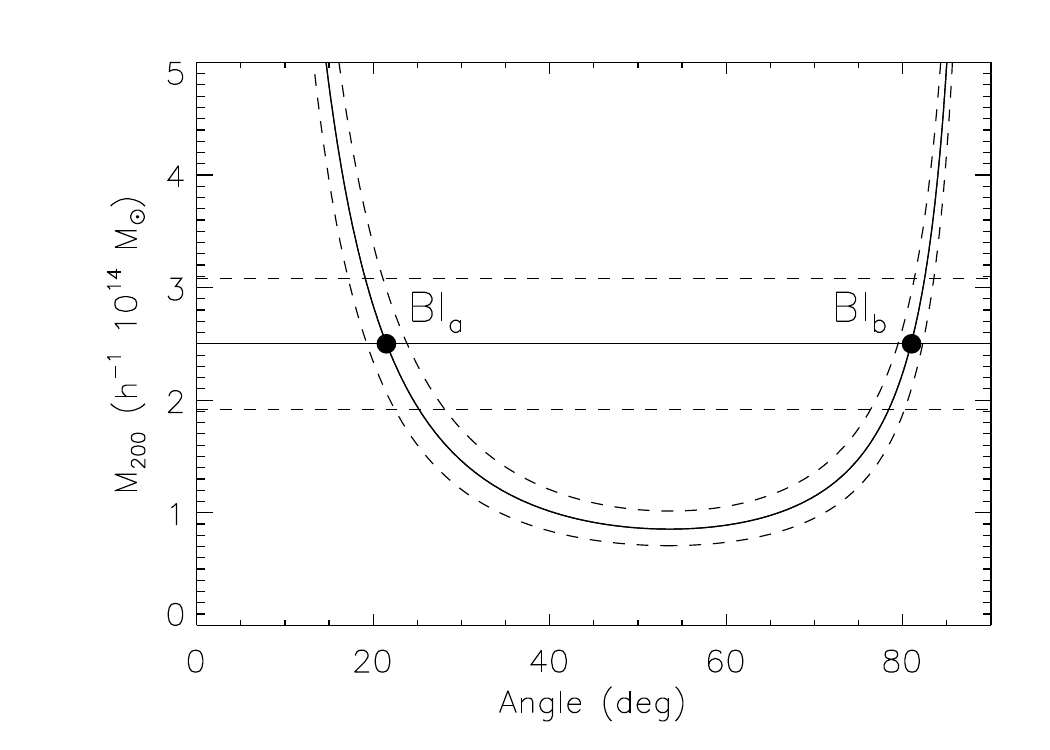}
\caption{Angle-mass representation for the two-body interaction model found for the main cluster 
body and substructure. Solid and dashed curves represent the bound incoming (BI) solution and 
uncertainties estimated as possible models with relative velocities 1400 and 1685 km s$^{-1}$. 
The total mass of the system is represented by the horizontal line, with its uncertainty (dashed 
lines). We find two possible bound incoming solutions (BI$_a$) and BI$_b$) at 22$^\circ$ and 
81$^\circ$ with respect to the plane of the sky.}
\label{fig:bimodal}
\end{figure}

The Newtonian criterion for the gravitational binding follows the expression $V_r^2 D \leq 2GM_{sys} 
\sin^2 \alpha \cos \alpha$, where $\alpha$ is the projection angle between the line connecting 
the centres of the two clumps and the plane of the sky. \citet{Bee82} and \citet{Tho82} 
\citep[see also][]{Lub98} developed the formalism but it suffers from several constraints: first, the 
components interact radially, so only head-on collisions with zero angular momentum (no rotation) 
are possible; Second, the evolution starts at $t_0=0$ with a separation of $d_0=0$, and the 
clumps are moving apart or coming together for the first time in the history in the model considered;
and third, the inequation above exposed implicitly excludes unbound solutions. In addition to these
limitations, we remark that the two clusters are treated as point masses, which is an assumption that 
clearly fails when the two cluster mass distributions overlap.

The solutions for this model are shown in Fig. \ref{fig:bimodal}, which shows a mass-angle representation 
of the model. Considering the value of M$_{sys}$, we only find bound and incoming solutions. No 
bound outgoing solutions are found. So the main cluster and the substructure would be completely tied 
by mutual pull of gravity. That is, we find that we are seeing the cluster in a first interaction (at $t=12.462$ 
Gyr at the redshift of RXCJ1111). Two possible bound solutions are found but they are degenerated due 
to the ambiguity in the projection angle $\alpha$. These are BI$_a$ and BI$_b$, which correspond to 
$22_{-3}^{+6}$ and $81_{-5}^{+2}$ degrees, respectively. Assuming for simplicity that BI$_a\sim 22^\circ$ 
and BI$_b\sim 81^\circ$, we estimate the actual de-projected (3D) relative distances and velocities between 
both clumps. In the first case, BI$_a\sim 22^\circ$, both galaxy clumps would be separated by a distance 
of $\Delta$d$_a=0.13$ Mpc, but the fossil substructure would be colliding with a relative 3D velocity of 
$\Delta \textrm{v}_a=4100$ km s$^{-1}$. In the second case, BI$_b\sim 81^\circ$, the main cluster and 
substructure would be separated by $\Delta$d$_b=0.70$ Mpc and would show a $\Delta \textrm{v}_b=1560$ km 
s$^{-1}$. Thus, in the first case, we find that the substructure would be very close to the main cluster, 
almost completely merged, as close as $0.1 r_{200}$, showing a very high speed, $>4000$ km s$^{-1}$. In 
the second scenario, the substructure would be separated $\sim 0.6 r_{200}$ from the main cluster, and 
colliding with a 3D velocity of $\sim 1560$ km s$^{-1}$.

The first scenario corresponds to a collision in a quite advanced stage, where halos and ICMs of both 
components would be almost completely fussed. With a so high relative velocity and both ICMs so mixed, 
we would expect to measure a high gas temperature, which is not observed in the X-ray maps. Nevertheless, 
the second scenario correspond to merging a whith a substructure at $\sim0.70$ Mpc from the main cluster 
centre with a still moderate velocity. This scenario is in agreement with an early stage merging, where 
ICMs are starting to interact, with a no so high X-ray temperature and no shock fronts observed. 
In fact, taking into account this geometry, we are viewing these systems from almost along 
the merger axis, in which case, if a shock front is present, we see it from within the Mach 
cone. In other words, we cannot see a sharp shock front because our line of sight is not tangent to 
the front. 

It is not easy to find a unique and satisfactory explanation for all the observables (BCG velocity, 
velocity distributions, velocity gradients, X-ray properties, spatial distributions, cluster morphology, 
...), and given the limitations of the Beers formalism, above exposed, it would be a mistake to 
completely rule out a more advanced merger on the basis of the Beers model. However, we find that the 
second scenario here proposed for the merging, represented by a bound incoming solution with a projection 
angle $\sim 81^\circ$ would be the preferred model to explain the merging state of RXCJ1111. In addition, 
we have to take into account that the substructure and \textit{BCG} show higher velocities with respect 
to the main cluster, thus the substructure is falling in from the front. At $t\sim0.15$ Gyr, the model 
predicts that the fossil substructure would be completely merged with the main cluster. 

\section{Summary and conclusions}
\label{sec:conclusions}

We present a detailed study of the kinematical and dynamical state of the galaxy cluster 
RXCJ1111+4050. Our analysis is based on new spectroscopic observations acquired at the 3.5m TNG 
telescope and complementary SDSS-DR16 spectroscopic redshifts in a region of $\sim 1r_{200}$. We
select 104 cluster members around $z=0.0756$. The study of the velocity field confirms the presence 
of significant deviations from Gaussianity, which have been explained by the presence of
a substructure in the cluster. The galaxy membership reveals that the secondary substructure is a
a fossil-like group, with a magnitude gap $\Delta m_{12}\sim 1.8$, merging with the main cluster.

X-ray surface brightness map shows a clear elongated shape in the North-South direction, which is 
in agreement with the 2D spatial distribution of galaxies and a velocity gradient of about 250-350 
km s$^{-1}$ Mpc$^{-1}$, also close to that direction. These facts, together with an indistinguishable 
galaxy populations projected onto the plane of the sky, indicate that the main cluster and 
substructure of RXCJ1111 lie almost aligned along the LOS, showing a small misalignment 
toward the south.

We use the velocity dispersion to estimate dynamical masses, and obtain M$_{200}=1.9 \pm 0.4 \times 
10^{14}$ M$_{\odot}$ and $0.6 \pm 0.4 \times 10^{14}$ M$_{\odot}$ for the main cluster and the 
secondary substructure, respectively. The total mass of RXCJ1111 derived from X-ray is in very 
good agreement with the dynamical estimates when individual galaxy clumps are considered, but not
when the cluster is assumed to be composed by a single component. This clearly suggests that
the most appropriated way to estimate dynamical masses in non-relaxed galaxy clusters is by
identifying galaxy clumps and computing the total mass as the sum of the different components, 
which should be more virialised than the whole cluster. In the end, the methodology followed here 
represents an example to obtain realistic dynamical masses, which is crucial, for instance,
to establish scaling relations from different approaches (X-ray, optical, weak lensing, ...). 

The observed excess in the X-ray temperature agrees with the fact that the substructure is starting 
to collide. This merger is characterized by a mass ratio of 3:1. We propose a possible 
merger model consistent with a two body configuration where cluster and substructure are aligned with 
$\sim 9^\circ (\pm 3^\circ)$ from the LOS, with an impact velocity of $\Delta v_{rf}\sim 1600$ 
km s$^{-1}$. This model also predicts that main cluster and fossil substructure will be completely 
joined in about 0.15 Gyr.

To summarise, RXCJ1111 represents an observational evidence that the fossil feature of galaxy 
systems is a transitional stage, which supports the results by \citet{vBB08} using 
simulations. The dynamical analysis here exposed demonstrates that a fossil-like group is 
falling into the RXCJ1111 main cluster. The on-going collision might accelerate the interaction 
between the three BCGs observed, making the cluster to show a smaller $\Delta m_{12}$ in the near 
future (in $<0.15$ Gyr), by fully incorporating \textit{BCG} and its corresponding galaxy population 
into the dynamics of the whole cluster and so losing its fossil condition.

\begin{acknowledgements}

R. Barrena acknowledges support by the Severo Ochoa 2020 research programme of 
the Instituto de Astrof\'{\i}sica de Canarias. G. Chon acknowledges support by 
the DLR under the grant n$^\circ$ 50 OR 2204. H. B\"ohringer acknowledges support 
from the Deutsche Forschungsgemeinschaft through the Excellence cluster 
"Origins". J.M.A. acknowledges the support of the Viera y Clavijo Senior program 
funded by ACIISI and ULL. J.M.A. acknowledges support from Spanish Ministerio de 
Ciencia, Innovaci\'on y Universidades through grant PID2021-128131NB-I00.\\

This article is based on observations made with the Italian Telescopio Nazionale 
Galileo operated by the Fundaci\'on Galileo Galilei of the INAF (Istituto Nazionale 
di Astrofisica). This facility is located at the Spanish del Roque de los 
Muchachos Observatory of the Instituto de Astrof\'{\i}sica de Canarias on the 
island of La Palma. \\

Funding for the Sloan Digital Sky Survey (SDSS) has been provided by the Alfred
P. Sloan Foundation, the Participating Institutions, the National Aeronautics
and Space Administration, the National Science Foundation, the U.S. Department
of Energy, the Japanese Monbukagakusho, and the Max Planck Society. 

\end{acknowledgements}

\bibliographystyle{aa}

\appendix
\section{Spectroscopic redshifts catalogue}

\begin{table}
\fontsize{9}{11}
\caption{Velocity catalogue in the RXJ1111 field considered in this work, which includes 109 new spectra observed in the 
3.5m TNG telescope and 43 complementary redshift obtained from SDSS DR-16 database.}
\label{tab:catalog}
\tiny
\begin{tabular}{l c c p{5mm} p{5mm} l}
\midrule \midrule
 ID & R.A. \& Dec. (J2000)              & v$\pm \Delta$v & \makebox[8mm][c]{$g^\prime$} & \makebox[8mm][c]{$r^\prime$} & Notes \cr
    & R.A.=$11\! : \! mm \! : \! ss.ss$ & (km s$^{-1}$)  &            &            & \cr
    & Dec.=$+40\! :\! mm \! : \! ss.s$  &                &            &            & \cr
\midrule
1$^\star$  & 10:43.04 \enspace 47:20.3 & 24888  $\pm$ 142 & 18.21 & 17.87 & \cr
2$^\star$  & 10:43.78 \enspace 51:21.9 & 22140  $\pm$ 92  & 19.31 & 18.49 & \cr
3          & 10:44.86 \enspace 51:15.0 & 45551  $\pm$ 81  & 19.43 & 18.88 & ELG \cr
4          & 10:46.61 \enspace 49:40.2 & 107820 $\pm$ 83  & 21.24 & 20.55 & ELG \cr
5$^\star$  & 10:47.91 \enspace 50:37.1 & 23595  $\pm$ 152 & 19.30 & 18.99 & ELG \cr
6 	   & 10:48.26 \enspace 51:44.3 & 79047  $\pm$ 78  & 19.52 & 18.30 & \cr
7 	   & 10:48.77 \enspace 47:46.2 & 58600  $\pm$ 117 & 19.51 & 19.05 & ELG \cr
8 	   & 10:52.98 \enspace 49:05.4 & 29410  $\pm$ 91  & 21.07 & 20.86 & ELG \cr
9$^\star$  & 10:52.99 \enspace 48:53.7 & 21300  $\pm$ 95  & 21.17 & 21.03 & ELG \cr
10	   & 10:57.82 \enspace 47:05.7 & 46174  $\pm$ 41  & 18.80 & 17.77 & ELG \cr
11	   & 10:58.58 \enspace 48:04.4 & 33869  $\pm$ 72  & 21.60 & 20.76 & \cr
12	   & 10:58.84 \enspace 46:55.5 & 55360  $\pm$ 105 & 20.65 & 20.24 & ELG \cr
13$^\star$ & 11:00.10 \enspace 47:21.4 & 22697  $\pm$ 6	& 17.79 & 16.97 & \cr
14	   & 11:04.03 \enspace 50:04.8 & 103930 $\pm$ 125 & 22.65 & 21.18 & ELG \cr
15	   & 11:04.35 \enspace 50:16.7 & 28460  $\pm$ 75  & 18.22 & 17.65 & ELG \cr
16	   & 11:04.63 \enspace 49:25.7 & 60454  $\pm$ 95  & 21.14 & 20.17 & \cr
17	   & 11:04.96 \enspace 49:54.8 & 52510  $\pm$ 75  & 20.53 & 19.91 & ELG \cr
18	   & 11:05.18 \enspace 50:59.3 & 45709  $\pm$ 64  & 19.13 & 18.57 & ELG \cr
19	   & 11:09.94 \enspace 52:03.2 & 51680  $\pm$ 109 & 20.91 & 20.58 & \cr
20$^\star$ & 11:11.40 \enspace 52:47.6 & 23826  $\pm$ 4	& 17.57 & 16.77 & \cr
21$^\star$ & 11:12.36 \enspace 57:34.3 & 21454  $\pm$ 5	& 17.34 & 16.47 & \cr
22	   & 11:12.53 \enspace 54:12.6 & 107261 $\pm$ 67  & 20.05 & 18.39 & \cr
23$^\star$ & 11:12.70 \enspace 52:24.5 & 22630  $\pm$ 7	& 18.12 & 17.21 & \cr
24$^\star$ & 11:17.24 \enspace 52:43.5 & 21620  $\pm$ 117 & 19.16 & 18.78 & ELG \cr
25$^\star$ & 11:19.66 \enspace 51:35.3 & 22398  $\pm$ 85  & 19.43 & 18.63 & \cr
26$^\star$ & 11:21.46 \enspace 55:03.6 & 22140  $\pm$ 7	& 17.59 & 16.73 & \cr
27$^\star$ & 11:23.10 \enspace 49:23.3 & 22241  $\pm$ 78  & 18.69 & 17.82 & \cr
28$^\star$ & 11:23.69 \enspace 49:44.3 & 22604  $\pm$ 10  & 18.51 & 17.66 & \cr
29	   & 11:24.51 \enspace 45:13.7 & 123028 $\pm$ 97  & 20.91 & 19.23 & \cr
30$^\star$ & 11:27.16 \enspace 55:07.8 & 23037  $\pm$ 6	& 17.33 & 16.35 & \cr
31$^\star$ & 11:27.24 \enspace 55:24.3 & 22096  $\pm$ 91  & 18.95 & 18.12 & \cr
32$^\star$ & 11:27.65 \enspace 56:48.4 & 23378  $\pm$ 5	& 17.05 & 16.14 & \cr
33$^\star$ & 11:28.17 \enspace 55:26.3 & 23096  $\pm$ 6	& 18.07 & 17.18 & \cr
34$^\star$ & 11:28.72 \enspace 57:27.1 & 22109  $\pm$ 12  & 17.65 & 16.77 & \cr
35$^\star$ & 11:30.25 \enspace 58:00.6 & 23586  $\pm$ 55  & 18.72 & 17.87 & \cr
36$^\star$ & 11:30.29 \enspace 51:33.0 & 22051  $\pm$ 2	& 17.95 & 17.56 & \cr
37	   & 11:31.94 \enspace 46:48.0 & 109534 $\pm$ 90  & 22.06 & 21.24 & \cr
38$^\star$ & 11:32.78 \enspace 53:55.8 & 21491  $\pm$ 116 & 20.64 & 19.40 & \cr
39$^\star$ & 11:32.88 \enspace 53:25.0 & 22290  $\pm$ 70  & 17.18 & 16.26 & \cr
40$^\star$ & 11:32.91 \enspace 53:25.0 & 22274  $\pm$ 33  & 17.12 & 16.22 & \cr
41	   & 11:33.87 \enspace 55:51.0 & 28459  $\pm$ 143 & 18.32 & 17.62 & ELG \cr
42$^\star$ & 11:34.05 \enspace 55:44.5 & 21046  $\pm$ 7	& 18.22 & 17.35 & \cr
43$^\star$ & 11:34.53 \enspace 47:12.4 & 22401  $\pm$ 63  & 18.24 & 17.50 & \cr
44$^\star$ & 11:35.07 \enspace 45:53.7 & 22037  $\pm$ 8	& 18.31 & 17.41 & \cr
45	   & 11:35.35 \enspace 41:43.5 & 154103 $\pm$ 76  & 21.53 & 19.96 & \cr
46$^\star$ & 11:35.45 \enspace 47:03.9 & 21734  $\pm$ 73  & 19.10 & 18.22 & \cr
47$^\star$ & 11:35.55 \enspace 57:49.3 & 21737  $\pm$ 72  & 18.67 & 17.81 & \cr
48$^\star$ & 11:35.77 \enspace 47:27.2 & 21888  $\pm$ 113 & 19.70 & 18.80 & \cr
49$^\star$ & 11:35.80 \enspace 42:26.3 & 22880  $\pm$ 5	& 17.95 & 17.08 & \cr
50$^\star$ & 11:36.17 \enspace 51:05.1 & 22510  $\pm$ 99  & 18.39 & 17.44 & \cr
51$^\star$ & 11:36.59 \enspace 51:21.5 & 20389  $\pm$ 90  & 18.25 & 17.42 & \cr
52$^\star$ & 11:36.63 \enspace 51:58.9 & 22633  $\pm$ 55  & 19.11 & 18.22 & \cr
53$^\star$ & 11:36.64 \enspace 42:06.7 & 22110  $\pm$ 83  & 20.85 & 20.60 & ELG \cr
54$^\star$ & 11:37.05 \enspace 50:29.5 & 23545  $\pm$ 7	& 17.73 & 16.82 & \cr
55	   & 11:38.06 \enspace 45:50.7 & 56993  $\pm$ 119 & 21.98 & 21.20 & \cr
56$^\star$ & 11:38.43 \enspace 44:20.4 & 22591  $\pm$ 117 & 19.94 & 19.13 & \cr
57$^\star$ & 11:38.68 \enspace 52:59.6 & 24479  $\pm$ 84  & 18.29 & 17.41 & \cr
58	   & 11:38.77 \enspace 56:55.6 & 34400  $\pm$ 105 & 18.97 & 18.68 & ELG \cr
59$^\star$ & 11:39.54 \enspace 51:14.9 & 22913  $\pm$ 140 & 20.33 & 19.48 & \cr
60$^\star$ & 11:39.73 \enspace 50:24.3 & 22283  $\pm$ 36  & 15.49 & 14.55 & \textit{BCG2} \cr
\midrule
\end{tabular}
\end{table}

\addtocounter{table}{-1}
\begin{table}
\fontsize{9}{11}
\caption{Continued.}
\tiny
\begin{tabular}{l c c p{5mm} p{5mm} l}
\midrule \midrule
 ID & R.A. \& Dec. (J2000)              & v$\pm \Delta$v & \makebox[8mm][c]{$g^\prime$} & \makebox[8mm][c]{$r^\prime$} & Notes \cr
    & R.A.=$11\! : \! mm \! : \! ss.ss$ & (km s$^{-1}$)  &            &            & \cr
    & Dec.=$+40\! :\! mm \! : \! ss.s$  &                &            &            & \cr
\midrule
61$^\star$ & 11:39.94 \enspace 58:16.5 & 23173  $\pm$ 13  & 18.37 & 17.60 & \cr
62$^\star$ & 11:39.98 \enspace 00:46.7 & 21729  $\pm$ 7       & 17.93 & 17.05 & \cr
63	 & 11:39.99 \enspace 53:13.7 & 41547  $\pm$ 95  & 20.15 & 19.17 & \cr
64$^\star$ & 11:40.09 \enspace 45:45.3 & 21669  $\pm$ 6       & 18.12 & 17.19 & \cr
65$^\star$ & 11:40.14 \enspace 44:44.0 & 24423  $\pm$ 60  & 19.08 & 18.18 & \cr
66$^\star$ & 11:40.20 \enspace 46:57.8 & 22503  $\pm$ 9       & 19.13 & 18.23 & \cr
67$^\star$ & 11:40.48 \enspace 47:06.9 & 22329  $\pm$ 5       & 16.15 & 15.21 & \textit{BCG3} \cr
68	 & 11:40.57 \enspace 44:01.7 & 105740 $\pm$ 119 & 21.53 & 20.64 & ELG \cr
69$^\star$ & 11:40.72 \enspace 50:02.9 & 24253  $\pm$ 192 & 18.85 & 17.99 & \cr
70	   & 11:40.75 \enspace 43:22.2 & 90605  $\pm$ 92  & 22.01 & 20.27 & \cr
71$^\star$ & 11:40.94 \enspace 50:31.0 & 21270  $\pm$ 67  & 19.48 & 18.59 & \cr
72$^\star$  & 11:41.00 \enspace 53:48.2 & 23981  $\pm$ 150 & 20.72 & 19.94 & \cr
73	    & 11:41.85 \enspace 46:26.6 & 81529  $\pm$ 46  & 20.49 & 19.01 & \cr
74	    & 11:41.96 \enspace 40:57.4 & 145220 $\pm$ 81  & 22.10 & 21.33 & ELG \cr
75$^\star$  & 11:42.39 \enspace 54:35.1 & 21750  $\pm$ 199 & 20.29 & 19.49 & \cr
76$^\star$  & 11:42.49 \enspace 45:10.7 & 21968  $\pm$ 5 & 16.78 & 15.88 & \cr
77$^\star$  & 11:42.89 \enspace 47:19.0 & 23074  $\pm$ 106 & 18.79 & 17.87 & \cr
78$^\star$  & 11:43.11 \enspace 43:33.9 & 22734  $\pm$ 31  & 17.38 & 16.46 & \cr
79	    & 11:43.13 \enspace 01:01.3 & 14610  $\pm$ 107 & 20.00 & 19.62 & ELG \cr
80$^\star$  & 11:43.23 \enspace 54:31.2 & 22331  $\pm$ 73  & 18.00 & 17.10 & \cr
81	    & 11:43.42 \enspace 46:38.7 & 86349  $\pm$ 65  & 20.95 & 20.58 & \cr
82$^\star$  & 11:43.48 \enspace 42:00.3 & 23794  $\pm$ 36  & 18.71 & 17.84 & \cr
83$^\star$  & 11:43.62 \enspace 49:14.5 & 23428  $\pm$ 5 & 15.11 & 14.17 & \textit{BCG} \cr
84$^\star$  & 11:43.71 \enspace 53:16.9 & 22854  $\pm$ 8 & 18.39 & 17.46 & \cr
85$^\star$  & 11:43.76 \enspace 50:05.3 & 21844  $\pm$ 8 & 18.58 & 17.68 & \cr
86$^\star$  & 11:43.81 \enspace 56:07.2 & 22871  $\pm$ 67  & 18.27 & 17.40 & \cr
87	    & 11:44.13 \enspace 46:07.4 & 124630 $\pm$ 145 & 21.53 & 19.93 & ELG \cr
88$^\star$  & 11:44.24 \enspace 48:29.1 & 24051  $\pm$ 60  & 19.40 & 18.43 & \cr
89	    & 11:44.55 \enspace 40:08.7 & 138350 $\pm$ 93  & 22.38 & 21.50 & ELG \cr
90	    & 11:44.87 \enspace 58:42.5 & 13890  $\pm$ 107 & 19.84 & 19.60 & ELG \cr
91$^\star$  & 11:44.93 \enspace 48:46.3 & 23202  $\pm$ 6 & 17.44 & 16.53 & \cr
92$^\star$  & 11:45.59 \enspace 45:28.7 & 22372  $\pm$ 123 & 21.54 & 20.66 & \cr
93$^\star$  & 11:45.65 \enspace 50:57.7 & 22798  $\pm$ 91  & 18.68 & 17.82 & \cr
94$^\star$  & 11:46.13 \enspace 51:19.6 & 23755  $\pm$ 7 & 18.38 & 17.45 & \cr
95	    & 11:46.22 \enspace 01:18.1 & 19159  $\pm$ 100 & 20.73 & 20.28 & \cr
96	    & 11:46.35 \enspace 59:24.7 & 70740  $\pm$ 78  & 20.33 & 19.13 & \cr
97$^\star$  & 11:47.19 \enspace 57:29.6 & 22287  $\pm$ 107 & 19.85 & 19.10 & \cr
98	    & 11:47.71 \enspace 54:13.1 & 45950  $\pm$ 100 & 19.74 & 19.15 & \cr
99$^\star$  & 11:48.54 \enspace 54:10.0 & 22998  $\pm$ 6	& 18.21 & 17.32 & \cr
100$^\star$ & 11:48.55 \enspace 49:42.2 & 23452  $\pm$ 106 & 19.96 & 19.11 & \cr
101$^\star$ & 11:49.37 \enspace 53:35.5 & 22650  $\pm$ 103 & 19.19 & 18.36 & \cr
102$^\star$ & 11:50.00 \enspace 52:26.8 & 22888  $\pm$ 114 & 19.89 & 18.99 & \cr
103$^\star$ & 11:50.35 \enspace 46:13.5 & 23679  $\pm$ 5	& 16.91 & 15.94 & \cr
104$^\star$ & 11:50.62 \enspace 51:21.0 & 23365  $\pm$ 5	& 17.50 & 16.58 & \cr
105$^\star$ & 11:50.85 \enspace 51:01.5 & 23093  $\pm$ 8	& 18.38 & 17.54 & \cr
106$^\star$ & 11:51.36 \enspace 54:36.5 & 22773  $\pm$ 6	& 17.41 & 16.51 & \cr
107	    & 11:51.37 \enspace 00:32.8 & 102542 $\pm$ 80  & 20.96 & 20.12 & ELG \cr
108$^\star$ & 11:51.46 \enspace 54:50.5 & 21871  $\pm$ 91  & 20.12 & 19.35 & \cr
109$^\star$ & 11:52.32 \enspace 52:36.9 & 23998  $\pm$ 69  & 19.71 & 18.88 & \cr
110$^\star$ & 11:52.32 \enspace 55:30.2 & 21548  $\pm$ 80  & 19.12 & 18.40 & \cr
111	    & 11:52.36 \enspace 53:06.3 & 70670  $\pm$ 55  & 21.28 & 20.60 & ELG \cr
112$^\star$ & 11:52.52 \enspace 58:32.8 & 24219  $\pm$ 9	& 18.12 & 17.24 & \cr
113	    & 11:54.22 \enspace 55:09.4 & 28475  $\pm$ 122 & 19.11 & 18.76 & ELG \cr
114$^\star$ & 11:54.23 \enspace 51:15.9 & 22657  $\pm$ 6	& 18.15 & 17.24 & \cr
115$^\star$ & 11:54.55 \enspace 45:47.5 & 23713  $\pm$ 4	& 18.22 & 17.31 & \cr
116	    & 11:54.59 \enspace 59:50.5 & 77484  $\pm$ 84  & 21.21 & 19.90 & \cr
117$^\star$ & 11:55.53 \enspace 47:19.3 & 22280  $\pm$ 78  & 18.55 & 17.85 & ELG \cr
118	    & 11:55.53 \enspace 47:19.6 & 21670  $\pm$ 103 & 18.49 & 17.81 & ELG \cr
119$^\star$ & 11:56.57 \enspace 51:42.3 & 23790  $\pm$ 7	& 17.62 & 16.74 & \cr
120$^\star$ & 11:57.15 \enspace 51:52.9 & 23297  $\pm$ 145 & 19.89 & 19.00 & \cr
121$^\star$ & 11:58.08 \enspace 48:26.8 & 23098  $\pm$ 7	& 18.06 & 17.17 & \cr
122	    & 11:58.18 \enspace 53:16.7 & 46207  $\pm$ 70  & 19.10 & 18.44 & ELG \cr
123$^\star$ & 11:58.80 \enspace 52:03.7 & 22343  $\pm$ 54  & 17.74 & 16.87 & \cr
124	    & 11:59.75 \enspace 49:00.2 & 35400  $\pm$ 97  & 20.26 & 19.43 & ELG \cr
125$^\star$ & 12:00.72 \enspace 48:29.3 & 24336  $\pm$ 96  & 19.32 & 18.48 & \cr
126$^\star$ & 12:02.15 \enspace 40:43.0 & 22686  $\pm$ 8	& 17.86 & 16.96 & \cr
127$^\star$ & 12:02.23 \enspace 48:52.3 & 22800  $\pm$ 100 & 20.57 & 19.93 & \cr
128$^\star$ & 12:03.99 \enspace 40:18.1 & 21845  $\pm$ 8	  & 18.46 & 17.65 & \cr
129	      & 12:04.35 \enspace 49:26.3 & 55573  $\pm$ 76  & 20.50 & 19.84 & \cr
130$^\star$ & 12:04.42 \enspace 52:49.6 & 24120  $\pm$ 111 & 20.65 & 20.18 & ELG \cr
131	      & 12:05.09 \enspace 50:57.4 & 43900  $\pm$ 91  & 20.23 & 19.62 & ELG \cr
132$^\star$ & 12:05.72 \enspace 52:43.5 & 22670  $\pm$ 34  & 18.65 & 17.82 & \cr
\midrule
\end{tabular}
\end{table}

\addtocounter{table}{-1}
\begin{table}
\fontsize{9}{11}
\caption{Continued.}
\tiny
\begin{tabular}{l c c p{5mm} p{5mm} l}
\midrule \midrule
 ID & R.A. \& Dec. (J2000)              & v$\pm \Delta$v & \makebox[8mm][c]{$g^\prime$} & \makebox[8mm][c]{$r^\prime$} & Notes \cr
    & R.A.=$11\! : \! mm \! : \! ss.ss$ & (km s$^{-1}$)  &            &            & \cr
    & Dec.=$+40\! :\! mm \! : \! ss.s$  &                &            &            & \cr
\midrule
133$^\star$ & 12:06.37 \enspace 52:25.5 & 22036  $\pm$ 93  & 19.97 & 19.22 & \cr
134$^\star$ & 12:07.42 \enspace 47:12.7 & 23170  $\pm$ 132 & 20.17 & 19.50 & ELG \cr
135$^\star$ & 12:07.76 \enspace 57:21.7 & 21126  $\pm$ 3      & 16.82 & 16.26 & \cr
136$^\star$ & 12:07.80 \enspace 56:54.9 & 21931  $\pm$ 3      & 17.86 & 16.94 & \cr
137$^\star$ & 12:08.35 \enspace 53:01.9 & 21430  $\pm$ 112 & 19.03 & 18.55 & ELG \cr
138$^\star$ & 12:08.41 \enspace 51:44.3 & 22540  $\pm$ 109 & 19.52 & 19.09 & ELG \cr
139	    & 12:09.17 \enspace 52:11.6 & 53810  $\pm$ 89  & 20.97 & 20.54 & ELG \cr
140$^\star$ & 12:10.87 \enspace 50:29.8 & 22594  $\pm$ 27  & 17.25 & 16.38 & \cr
141	    & 12:12.61 \enspace 48:17.5 & 46487  $\pm$ 67  & 19.97 & 19.16 & ELG \cr
142$^\star$ & 12:12.87 \enspace 51:07.2 & 22074  $\pm$ 49  & 19.33 & 18.50 & \cr
143	    & 12:13.20 \enspace 51:36.0 & 41490  $\pm$ 68  & 19.98 & 19.60 & ELG \cr
144$^\star$ & 12:17.46 \enspace 52:42.8 & 22191  $\pm$ 5	& 17.38 & 16.48 & \cr
145$^\star$ & 12:18.01 \enspace 41:40.4 & 21502  $\pm$ 7	& 18.25 & 17.38 & \cr
146$^\star$ & 12:19.87 \enspace 46:34.2 & 24397  $\pm$ 71  & 20.29 & 19.59 & \cr
147	    & 12:20.32 \enspace 47:56.7 & 59000  $\pm$ 88  & 21.99 & 20.69 & \cr
148$^\star$ & 12:24.10 \enspace 48:37.8 & 22895  $\pm$ 114 & 19.82 & 19.06 & \cr
149$^\star$ & 12:25.98 \enspace 57:46.4 & 21461  $\pm$ 6	& 17.95 & 17.16 & \cr
150	    & 12:30.98 \enspace 54:32.2 & 45420  $\pm$ 115 & 19.47 & 18.95 & ELG \cr
151	    & 12:32.39 \enspace 54:06.4 & 45750  $\pm$ 80  & 20.60 & 20.19 & ELG \cr
152$^\star$ & 12:34.82 \enspace 49:41.8 & 22214  $\pm$ 44  & 18.89 & 18.04 & \cr
\midrule
\end{tabular}
\footnotesize{Note: asterisk in Col. 1 (ID) indicates the galaxies selected as cluster members.}
\end{table}


\end{document}